\PassOptionsToPackage{usenames, dvipsnames, svgnames, table}{xcolor}
\documentclass[letterpaper,twocolumn,10pt]{article}
\usepackage{usenix-2020-09}

\newif \ifcomments
\commentstrue

\newif \iffullversion
\ifdefined\isfull
\fullversiontrue
\else
\fullversionfalse
\fi
\fullversiontrue

\usepackage{times}
\usepackage{bbm}
\usepackage{varwidth}
\usepackage{xspace}
\usepackage{multirow}
\usepackage{comment}
\usepackage{xfrac}
\usepackage{float}
\usepackage{booktabs} 
\usepackage{fancybox, fancyvrb, calc}
\usepackage[center, tight]{subfigure}
\usepackage[inline]{enumitem}
\usepackage{amsmath, amssymb}
\usepackage[font=small]{caption}
\usepackage{url}
\usepackage{setspace}
\usepackage{breakurl}
\usepackage{graphicx}
\usepackage{pbox}
\usepackage{algorithm, algorithmicx}
\usepackage[noend]{algpseudocode}
\usepackage{amsthm}
\usepackage{balance}
\usepackage{wrapfig}
\usepackage{tikz, pgfplots}
\pgfplotsset{compat=newest}
\usetikzlibrary{patterns,matrix,positioning,shadows,backgrounds,arrows,calc,fit,automata,shapes.geometric,arrows,decorations.pathreplacing,decorations.markings}
\usepgfplotslibrary{groupplots}
\usepgfplotslibrary{dateplot}
\usepackage{tabularx}
\usepackage[T1]{fontenc}
\usepackage[utf8]{inputenc}
\usepackage[normalem]{ulem}
\usepackage{wasysym}
\usepackage{enumitem}
\usepackage{sidecap}
\usepackage{caption}
\usepackage{units}
\usepackage{pifont}
\usepackage[mathcal]{euscript}

\usepackage[noindentafter]{titlesec}

\titlespacing\section{0pt}{4pt plus 2pt minus 2pt}{4pt plus 2pt minus 2pt}
\titlespacing\subsection{0pt}{4pt plus 2pt minus 2pt}{4pt plus 2pt minus 2pt}
\usepackage{hyperref}
\hypersetup{
  colorlinks=true,      
  linkcolor=blue,       
  citecolor=magenta,    
  filecolor=cyan,       
  urlcolor=blue          
}
\usepackage[capitalise]{cleveref}

\newtheorem{theorem}{Theorem}

\newtheorem{invariant}{Invariant}

\definecolor{boxcolor}{gray}{0.9}

\makeatletter
\newcommand{\thickhline}{%
    \noalign {\ifnum 0=`}\fi \hrule height 0.8pt
    \futurelet \reserved@a \@xhline
}
\newcolumntype{"}{@{\vrule width 0.8pt}}
\newcolumntype{[}{@{\vrule width 0.8pt\hskip\tabcolsep}}
\newcolumntype{]}{@{\hskip\tabcolsep\vrule width 0.8pt}}
\newcolumntype{!}{@{\hskip\tabcolsep\vrule width 0.8pt\hskip\tabcolsep}}
\makeatother

\newcommand{\bnm}{\begin{newmath}}
\newcommand{\enm}{\end{newmath}}

\newcommand{\bea}{\begin{eqnarray*}}
\newcommand{\eea}{\end{eqnarray*}}

\newcommand{\bne}{\begin{newequation}}
\newcommand{\ene}{\end{newequation}}

\newenvironment{newmath}{\begin{displaymath}%
\setlength{\abovedisplayskip}{4pt}%
\setlength{\belowdisplayskip}{4pt}%
\setlength{\abovedisplayshortskip}{6pt}%
\setlength{\belowdisplayshortskip}{6pt} }{\end{displaymath}}

\newenvironment{newequation}{\begin{equation}%
\setlength{\abovedisplayskip}{4pt}%
\setlength{\belowdisplayskip}{4pt}%
\setlength{\abovedisplayshortskip}{6pt}%
\setlength{\belowdisplayshortskip}{6pt} }{\end{equation}}

\algdef{SE}[DOWHILE]{Do}{doWhile}{\algorithmicdo}[1]{\algorithmicwhile\ #1}%

\definecolor{britishracinggreen}{rgb}{0.0, 0.26, 0.15}
\definecolor{ao(english)}{rgb}{0.0, 0.5, 0.0}

\newcommand{\cut}[1]{}

\newcommand{\paragraphb}[1]{\vspace{0.075in}\noindent{\bf #1.}}

\newcommand{\paragraphc}[1]{\vspace{0.075in}\noindent{\em #1}}

\cut{
\usepackage{draftwatermark}
\SetWatermarkText{DRAFT -- DO NOT REDISTRIBUTE \ \ \ \ \ \ \ DRAFT -- DO NOT REDISTRIBUTE}
\SetWatermarkFontSize{1.5cm}
\SetWatermarkAngle{307}
\SetWatermarkLightness{0.85}
}

\definecolor{darkgreen}{rgb}{0.1, 0.14, 0.13}

\tikzset{ 
table/.style={
  matrix of nodes,
  row sep=-\pgflinewidth,
  column sep=-\pgflinewidth,
  nodes={rectangle,thick,draw=black,text width={},align=center,font=\small},
  text depth=0.25ex,
  text height=1.25ex,
  nodes in empty cells
},
map/.style={
  matrix of nodes,
  row sep=-\pgflinewidth,
  column sep=-\pgflinewidth,
  nodes={rectangle,draw=black,text width=5em,align=center,font=\small},
  text depth=0.25ex,
  text height=1.25ex,
  nodes in empty cells
},
bigmap/.style={
  matrix of nodes,
  row sep=-\pgflinewidth,
  column sep=-\pgflinewidth,
  nodes={rectangle,draw=black,text width=26em,align=center,font=\small},
  text depth=0.25ex,
  text height=1.25ex,
  nodes in empty cells
},
}

\tikzset{three sided/.style={
        draw=none,
        append after command={
            [shorten <= -0.5\pgflinewidth]
            ([shift={(-1.5\pgflinewidth,-0.5\pgflinewidth)}]\tikzlastnode.north west)
        edge([shift={( 0.5\pgflinewidth,-0.5\pgflinewidth)}]\tikzlastnode.north east) 
            ([shift={( 0.5\pgflinewidth,-0.5\pgflinewidth)}]\tikzlastnode.north east)
        edge([shift={( 0.5\pgflinewidth,+0.5\pgflinewidth)}]\tikzlastnode.south east)            
            ([shift={( 0.5\pgflinewidth,+0.5\pgflinewidth)}]\tikzlastnode.south east)
        edge([shift={(-1.0\pgflinewidth,+0.5\pgflinewidth)}]\tikzlastnode.south west)
        }
    }
}

\tikzstyle{startstop} = [rectangle, rounded corners, minimum width=3em, minimum height=1em,text centered, draw=black, fill=red!30]
\tikzstyle{io} = [trapezium, trapezium left angle=70, trapezium right angle=120, minimum width=2.5em, minimum height=1em, text centered, draw=black, fill=blue!30]
\tikzstyle{process} = [rectangle, minimum width=1.5em, minimum height=1em, align=center, draw=black, fill=gray!30]
\tikzstyle{decision} = [diamond, minimum width=3em, minimum height=1em, align=center, draw=black, fill=SkyBlue!30]
\tikzstyle{arrow} = [thick,->,>=stealth]
\tikzstyle{monolog} = [fill=SkyBlue!30]

\tikzset{
  on each segment/.style={
    decorate,
    decoration={
      show path construction,
      moveto code={},
      lineto code={
        \path [#1]
        (\tikzinputsegmentfirst) -- (\tikzinputsegmentlast);
      },
      curveto code={
        \path [#1] (\tikzinputsegmentfirst)
        .. controls
        (\tikzinputsegmentsupporta) and (\tikzinputsegmentsupportb)
        ..
        (\tikzinputsegmentlast);
      },
      closepath code={
        \path [#1]
        (\tikzinputsegmentfirst) -- (\tikzinputsegmentlast);
      },
    },
  },
  mid arrow/.style={postaction={decorate,decoration={
        markings,
        mark=at position .5 with {\arrow[#1]{stealth}}
      }}},
}

\algnewcommand{\IIf}[1]{\State\algorithmicif\ #1\ \algorithmicthen}%
\algnewcommand{\EndIIf}{\unskip\ }%
\algdef{SE}[DOWHILE]{Do}{doWhile}{\algorithmicdo}[1]{\algorithmicwhile\ #1}%
\algnewcommand\algorithmicforeach{\textbf{for each}}%
\algdef{S}[FOR]{ForEach}[1]{\algorithmicforeach\ #1\ \algorithmicdo}%

\newcommand{\gamesfontsize}{\footnotesize}
\newcommand{\fpage}[2]{\framebox{\begin{minipage}[t]{#1\textwidth}\setstretch{1.05}\gamesfontsize  #2 \end{minipage}}}

\newif\ifFullVersion
\FullVersionfalse

\def\ie{{i.e.}}
\def\eg{{\em e.g.}\xspace}

\def\name{\textsc{Shortstack}\xspace}
\def\pancake{\textsc{Pancake}\xspace}

\ifcomments
    \newcommand{\midhul}[1]{{\color{OrangeRed}{[Midhul: #1]}}}
    \newcommand{\kushal}[1]{{\color{Magenta}{[Kushal: #1]}}}
    \newcommand{\anurag}[1]{{\color{Purple}{[Anurag: #1]}}}
    \newcommand{\rachit}[1]{{\color{OrangeRed}{[Rachit: #1]}}}
    \newcommand{\reviewer}[2]{{\color{Red}{[Reviewer #1: #2]}}}
    \newcommand{\addressed}[2]{{\color{blue}{[Reviewer #1: #2]}}}
\else
    \newcommand{\midhul}[1]{}
    \newcommand{\kushal}[1]{}
    \newcommand{\anurag}[1]{}
    \newcommand{\rachit}[1]{}
    \newcommand{\reviewer}[2]{}
    \newcommand{\addressed}[2]{}
\fi
\newcommand{\maybeaddressed}[2]{}

\newcommand{\ninj}{\mathcal{P}}

\newcommand{\dist}{\pi}
\newcommand{\estdist}{\hat{\pi}}
\newcommand{\DB}{\mathsf{KV}}

\newcommand{\KeyGen}{\mathsf{KeyGen}}
\newcommand{\Enc}{\mathsf{Enc}}
\newcommand{\Dec}{\mathsf{Dec}}

\newcommand{\st}{\mathsf{st}}

\newcommand{\Init}{\mathsf{Init}}

\newcommand{\Process}{\mathsf{Process}\xspace}
\newcommand{\ProcessRequest}{\mathsf{ProcessQuery}\xspace}

\newcommand{\Weights}{\mathsf{Weights}}
\newcommand{\Configure}{\mathsf{Configure}}

\newcommand{\EDB}{\mathsf{EKV}}

\newcommand{\getsr}{{\:{\leftarrow{\hspace*{-3pt}\raisebox{.75pt}{$\scriptscriptstyle\$$}}}\:}}
\newcommand{\sample}{\getsr}

\newcommand{\getsbiased}[1]{{\:{\leftarrow{\hspace*{-3pt}\raisebox{.75pt}{$\scriptscriptstyle#1$}}}\:}}

\newcommand{\kw}{k}
\newcommand{\val}{v}

\newcommand{\ind}{\myind}

\newcommand{\procedurev}[1]{\underline{{#1}:}\smallskip}
\newcommand{\numqueries}{q}

\newcommand{\advA}{\mathbb{A}}
\newcommand{\adv}{\advA}
\newcommand{\advB}{\mathbb{B}}
\newcommand{\bdv}{\advB}
\newcommand{\advC}{\mathbb{C}}
\newcommand{\cdv}{\advC}
\newcommand{\advD}{\mathbb{D}}
\newcommand{\ddv}{\advD}
\newcommand{\myind}{\hspace*{1em}}

\newcommand{\AdvPRF}{\mathbf{Adv}^{\text{prf}}}
\newcommand{\AdvROR}{\mathbf{Adv}^{\text{ror}}}

\newcommand{\AdvDIST}{\mathbf{Adv}^{\text{dist}}}

\newcommand{\Batch}{\text{Batch}}

\newcommand{\Queue}{\mathsf{Queue}}

\newcommand{\chains}{\mathcal{C}}
\newcommand{\Retries}{\rho}
\newcommand{\failures}{\mathcal{T}}
\newcommand{\Interleave}{\mathsf{Interleave}}
\newcommand{\stadv}{\st_A}

\newcommand{\Transform}{\mathsf{Transform}}

\newcommand{\Shuffle}{\mathsf{Shuffle}}

\newcommand{\fakedist}{\ensuremath{\dist_{f}}}

\newcommand{\newdist}{\ensuremath{\dist^{\prime}}}
\newcommand{\newestdist}{\ensuremath{\estdist^{\prime}}}

\newcommand{\Pop}{\mathsf{Pop}}

\newcommand{\Empty}{\mathsf{Empty}}

\newcommand{\updatecache}{\ensuremath{\mathsf{UpdateCache}}\xspace}

\newcounter{mynote}[section]

\newcommand{\INDCDA}{\textsc{ind-cdfa}}
\newcommand{\INDCDDA}{\textsc{ind-cddfa}}

\newcommand{\AdvINDCDA}{\mathbf{Adv}^{\text{ind-cdfa}}}
\newcommand{\AdvINDCDDA}{\mathbf{Adv}^{\text{ind-cddfa}}}

\newcommand{\ROR}{\textnormal{ROR}\xspace}
\newcommand{\IND}{\textnormal{IND}\xspace}

\renewcommand{\paragraph}[1]{\vspace*{4pt}\noindent\textbf{#1}}

\newlength{\saveparindent}
\newlength{\saveparskip}
\newcounter{ctr}

\newcommand{\li}{\texttt{L1}\xspace}
\newcommand{\lii}{\texttt{L2}\xspace}
\newcommand{\liii}{\texttt{L3}\xspace}

\usepackage{graphicx}
\usepackage{epstopdf}
\usepackage[export]{adjustbox}
\usepackage{etoolbox}
\makeatletter
\patchcmd{\maketitle}
	{\@maketitle}
	{\@maketitle\vspace{-1em}}
	{}
	{}
\makeatother

\usepackage{printlen}
\usepackage[available,functional,reproduced]{usenixbadges}

\usepackage[symbol]{footmisc}

\makeatletter
\newcommand\thefontsize{The current font size is: \f@size pt}
\makeatother

\begin{document}

\date{}

\title{\Large \bf \name: Distributed, Fault-tolerant, Oblivious Data Access}

\author{
{\rm Midhul Vuppalapati\thanks{Equal contributions.}}
\\{\rm Cornell University}
\and
{\rm Kushal Babel\footnotemark[1]}
\\{\rm Cornell University}
\and
{\rm Anurag Khandelwal}
\\{\rm Yale University}
\and
{\rm Rachit Agarwal}
\\{\rm Cornell University}
}

\maketitle




\begin{abstract}
Many applications that benefit from data offload to cloud services operate on private data. A now-long line of work has shown that, even when data is offloaded in an encrypted form, an adversary can learn sensitive information by analyzing data access patterns. Existing techniques for oblivious data access---that protect against access pattern attacks---require a centralized and stateful trusted proxy to orchestrate data accesses from applications to cloud services. We show that, in failure-prone deployments, such a centralized and stateful proxy results in violation of oblivious data access security guarantees and/or in system unavailability. We thus initiate the study of distributed, fault-tolerant, oblivious data access. 

We present \name, a distributed proxy architecture for oblivious data access in failure-prone deployments. \name achieves the classical obliviousness guarantee---access patterns observed by the adversary being independent of the input---even under a powerful passive persistent adversary that can force failure of arbitrary (bounded-sized) subset of proxy servers at arbitrary times. We also introduce a security model that enables studying oblivious data access with distributed, failure-prone, servers. We provide a formal proof that \name enables oblivious data access under this model, and show empirically that \name performance scales near-linearly with number of distributed proxy servers.
\end{abstract}

\begin{sloppypar}
\section{Introduction}
\label{sec:intro}
Cloud services offer applications scalable, fault-tolerant, and easy-to-manage systems for storing and querying data. Many applications that benefit from offloading data to these cloud services operate on private data that can reveal sensitive information even when stored in an encrypted form~\cite{islam2012access,cash2015leakage,kellaris2016generic,grubbs2019learning,kornaropoulos2019data,pancake}. An example is that of medical practices offloading patient health data to the cloud~\cite{ehr1, kuo2011opportunities, microsoft-europe-medical}---charts accessed by oncologists can reveal not only whether a patient has cancer, but also depending on the frequency of accesses (\eg, the frequency of chemotherapy appointments), indicate the cancer's type and severity. Several such applications are subject to severe security concerns.

There is a large and active body of research on building systems for oblivious data access, that is, hiding not only the content of the data, but also data access patterns (\eg, access frequency across data objects). These systems use one of the two techniques---Oblivious RAM~\cite{oram, obladi, privatefs, oblivistore, shroud, taostore, curious, concuroram} that enables oblivious data access against active adversaries but has bandwidth overheads that are logarithmic in the number of data objects, or Pancake~\cite{pancake, mavroforakis2015modular, frequencysmoothing} that enables oblivious data access against passive persistent adversaries with a small constant bandwidth overhead. Both of these techniques provide a powerful oblivious data access guarantee: an adversary observing all queries to and all responses from the cloud storage service observes uniform random accesses over the encrypted data objects. The challenge, however, is that both of these techniques require a centralized, {\em stateful}, proxy to orchestrate data access from applications to cloud services. Such a centralized and stateful proxy means that existing systems for oblivious data access suffer from two core issues (\S\ref{ssec:beefy-single-proxy}): 

\vspace{-0.5em}
\begin{itemize}[itemsep=0pt, leftmargin=*]
  \item {\em Security violation, or long periods of system unavailability during proxy failures:} The proxy being stateful means that, upon a failure, the proxy may lose state. We show in \S\ref{ssec:beefy-single-proxy} that, if the proxy state is lost, na\"ively restarting a new proxy and executing queries without restoring the state would lead to violation of oblivious data access security guarantees. To avoid such a security violation, upon restarting a new proxy, the state must be restored before executing any queries, \eg, by downloading the entire data and metadata from the cloud, decrypting all the data, reconstructing the (ORAM or Pancake) data structure, re-encrypting all the data, and uploading all the data back to the storage service; this would lead to long periods of system unavailability.
  \item {\em Bandwidth and/or compute scalability bottlenecks:} Since the proxy receives multiple responses for each client query, it has bandwidth overheads ($\Omega(\log n)$ in ORAM~\cite{boyle2016there,larsen2018yes,persiano2019lower,weiss2018there,larsen2019lower,patel2019what} and $3\times$ in Pancake~\cite{pancake}); and, since the proxy is responsible for both data encryption/decryption and processing for each individual query and response, it has non-trivial compute overheads. Thus, the centralized proxy can become bandwidth or compute bottlenecked, limiting system throughput. 
\end{itemize}

\vspace{-0.5em}
\noindent
We present \name, a distributed, fault-tolerant, system for oblivious data access. \name achieves three desirable goals: (1) formal oblivious data access guarantee against passive persistent adversaries, even under failures; (2) system availability even when an arbitrary, bounded-sized, subset of distributed proxy servers may fail; and (3) near-linear throughput scalability with number of distributed proxy servers. In designing \name, we make three core contributions.

Our first contribution is to fundamentally establish security goals for oblivious data access in failure-prone deployments. Indeed, existing security models~\cite{oram, obladi, privatefs, oblivistore, shroud, taostore, curious, concuroram, pancake, dauterman2021snoopy} do not capture failures. We introduce a formal security model and a security definition to study distributed, fault-tolerant, systems for oblivious data access under passive persistent adversaries. The model requires the classical oblivious data access guarantee~\cite{oram, pancake}---access patterns observed by the adversary must be independent of the input; in addition, to capture failures, the model requires this guarantee to hold under a powerful adversary that can fail an arbitrary (bounded-sized) subset of distributed proxy servers at arbitrary times. Informally, under our security definition, a scheme is considered secure if the access distribution over encrypted data objects is independent of the input distribution, even with adversarial choice and time of proxy server failures. 

\begin{figure}
    \centering
    \includegraphics[width=\linewidth]{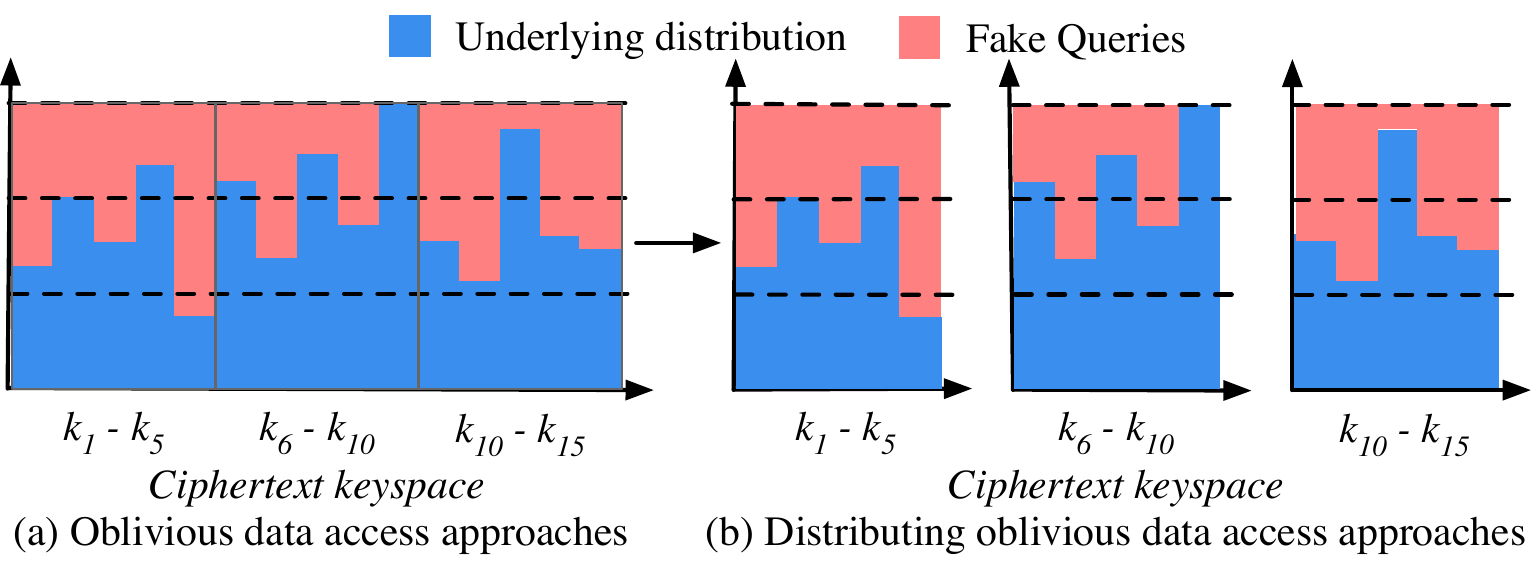}
    \caption{The flattened distribution over all ciphertext keys in oblivious data access schemes can be expressed as a sum of distributions over disjoint subsets of ciphertext keys.
    }
    \label{fig:sumofdist}
    \vspace{-1.5em}
\end{figure}

Our second contribution is design of a distributed, fault-tolerant, proxy architecture---\name---that enables oblivious data access against passive persistent adversaries, system availability (under a bounded number of failures), and near-linear throughput scalability with number of proxy servers. Simultaneously guaranteeing these three properties, especially when proxy servers can fail, turns out to be challenging: to avoid bandwidth and compute bottlenecks, multiple proxy servers must simultaneously process and send queries to the storage server; this makes it non-trivial, if not impossible, to ensure uniform random access over encrypted objects at all times (\eg, right after one of the proxy server fails) without giving up on availability. The key insight in \name design is that obliviousness only necessitates that access patterns observed by the adversary are independent of the input; the requirement of uniform random access over all encrypted objects as in prior designs is one, but not the only, way to achieve such independence. \name design achieves such independence as follows. The security of oblivious data access techniques stems from ``flattening'' the access distribution over unencrypted (plaintext) objects to a uniform random one over encrypted (ciphertext) objects (Figure~\ref{fig:sumofdist}~(a)). As illustrated visually in Figure~\ref{fig:sumofdist}~(b), any uniform random distribution over ciphertext objects can be decomposed into multiple sub-distributions in a manner that (1) each sub-distribution is uniform random over its support; and (2) the set of objects in any sub-distribution is equal in size, disjoint, and random. Thus, if each proxy server that forwards queries to the storage server is responsible for one of the sub-distributions, even with failure of a subset of these proxy servers, the adversary observes nothing but a uniform distribution (using (1)) over a random subset (using (2)) of objects. Achieving independence, and not necessarily a uniform random access pattern, at all times is at the core of the \name design. In \S\ref{sec:design}, we present a novel layered \name architecture that, using $k$ physical proxy servers, maintains system availability with up to $(k-1)$ proxy server failures and achieves throughput a factor ${\sim}k$ higher than a single proxy, all while enabling oblivious data access.

Our third contribution is a formal proof that \name enables oblivious data access under the above security model, and empirical evidence that \name can achieve near-perfect scalability with number of proxy servers (assuming storage server is not the bottleneck). We also show that \name gracefully handles failures: in the worst-case, \name throughput reduces linearly with number of proxy server failures (as one would expect). For the current \name implementation, the cost of achieving oblivious data access, availability and scalability is a ${\sim}7$ms increase in latency, a tiny fraction of the usual wide-area network latency. An end-to-end implementation
of \name is available at \url{https://github.com/pancake-security/shortstack}.

\section{\name Background}
\label{sec:overview}
We describe our system, failure, and threat models, followed by a brief primer on oblivious data access approaches.

\subsection{System, Threat and Failure Models}
\label{ssec:prelim}

\paragraphb{System model}
We consider settings where applications offload data to the cloud to benefit from the many properties enabled by cloud services, \eg, strong data durability and persistence, geo-replication, lower cost than provisioning dedicated and replicated storage servers, transparent handling of devices wearing out, and others. Examples of such applications include cloud-based healthcare services~\cite{europa-health,microsoft-europe-medical,euractive-microsoft,microsoft-eu-localization} as well as classical applications from access pattern attack literature~\cite{obladi, pancake}. The cloud-based storage service implements a key-value (KV) store that stores a collection of KV pairs, and support the following single-key operations: get, put, and delete. \name design can be applied to any data store that supports single-key read/write/delete operations. 

\name employs the standard encryption proxy model, commonly used in encrypted data stores~\cite{pancake, curious, taostore, cryptdb, baffle,ciphercloud,navajo,perspecsys,shn}: a trusted proxy orchestrates query execution from one or more client applications; the only difference compared to previous designs is that, in \name architecture, the proxy is logically-centralized but physically-distributed---that is, client queries may now be routed though multiple physical proxy servers within the same trusted domain. 

All network channels are encrypted using TLS. Each key $k$ in the KV store is encrypted using a pseudorandom function (PRF), denoted by $F(k)$; each value $v$ is symmetrically encrypted, denoted by $E(v)$. The logically-centralized proxy stores secret cryptographic keys needed for $F$ and $E$, and performs encryption. Since $F$ is deterministic, the proxy can execute all queries related to key $k$ by sending $F(k)$ to the cloud service. Similar to many existing commercial deployments~\cite{baffle,ciphercloud,navajo,perspecsys,shn}, keys and values are padded to a fixed size to avoid any length-based leakage.

\paragraphb{Threat model}
\name builds upon the widely-used trusted proxy threat model~\cite{taostore, oblivistore, privatefs, obladi}, where one or more mutually-trusting clients execute operations on an untrusted cloud storage service via a trusted proxy; as mentioned earlier, the only difference in \name is that the proxy is logically-centralized but comprises physically-distributed servers. As in many prior works~\cite{cryptdb, pancake, obladi}, we consider scenarios where the clients and the proxy servers all belong to a trusted domain. The storage service is controlled by an honest-but-curious (or, a passive persistent) adversary that observes all encrypted accesses but does not actively perform its own accesses. Since network channels are encrypted using TLS, the adversary cannot observe communications within the trusted domain, that is, the adversary cannot observe traffic between the clients and proxy servers. 

We model queries to the KV store using the Pancake model~\cite{pancake}: queries are generated as a sequence of accesses sampled from a (time-varying) distribution $\pi$ over $n$ KV pairs. While the encryption mechanism has  an estimate of the distribution $\hat\pi$, the adversary knows both the distribution $\pi$ and the transcript of encrypted queries and responses. We define a formal security model and definition in \S\ref{sec:security}, but informally, the system is secure if the transcript is \emph{independent} of the underlying distribution $\pi$, \ie, the adversary cannot identify an association between the two. 

\paragraphb{Failure model} We assume the cloud service provides data durability. However, proxy servers can fail. We consider the fail-stop model~\cite{fail-stop} for proxy server failures. 

\subsection{Oblivious Data Access Approaches}
\label{ssec:noise}
There are two approaches to oblivious data access today---the classical ORAM~\cite{oram, obladi, privatefs, oblivistore, shroud, taostore, curious, concuroram}, and the more recent approach of frequency smoothing as in Pancake~\cite{pancake, mavroforakis2015modular, frequencysmoothing}. ORAMs are designed to prevent a broad range of attacks (\eg, active adversaries); accordingly, they also suffer from high overheads, \eg, recent results~\cite{boyle2016there,larsen2018yes,persiano2019lower,weiss2018there,larsen2019lower,patel2019what} have established strong lower bounds on ORAM overheads---for a data store with $n$ KV pairs, any ORAM design must incur bandwidth overheads of $\Omega(\log n)$ (for proxy storage sublinear in KV store size). For KV stores that store millions or billions of KV pairs, these overheads may amount to orders-of-magnitude of throughput loss~\cite{pancake, oram-vldb}, making ORAMs impractical. Pancake enables oblivious data access against passive persistent adversaries, and incurs a small, constant, bandwidth overhead of $3\times$, independent of the number of objects in the KV store. Thus, we focus on building a distributed, fault-tolerant, proxy architecture within the Pancake context. To keep the paper self-contained, we summarize the Pancake mechanisms necessary to understand the \name architecture.

\begin{figure}
    \centering
    \includegraphics[width=0.8\linewidth]{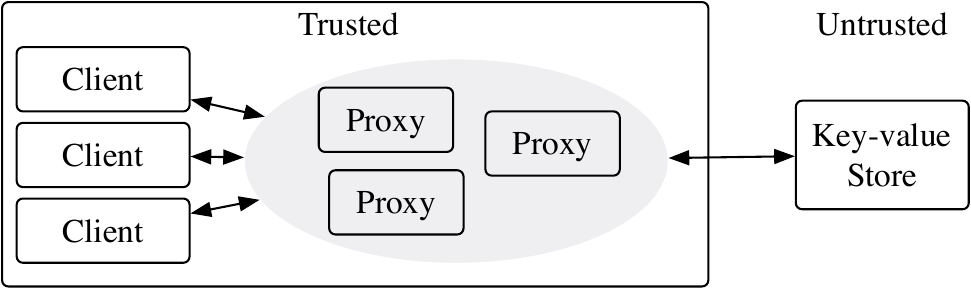}
    \caption{\name System and Threat model}
    \label{fig:fastpancakesetup}
    \vspace{-1.5em}
\end{figure}

\paragraphb{A brief primer to oblivious data access using Pancake} 
The Pancake approach combines the knowledge of the distribution estimate $\hat\pi$ with several techniques (selective replication, fake accesses, batching, etc.) to transform a sequence of queries into uniform accesses over encrypted KV pairs.
Selective replication creates ``replicas'' of KV pairs that 
have high access probability relative to other KV pairs, which serves to partially 
smooth the distribution over (replicated) KV pairs, while also ensuring that the total number of 
keys to be stored in the KV store is exactly $2n$ (if needed, dummy replicas are added so that the number of replicas does not reveal any distribution-sensitive information). To hide the association between the original keys and their replicas, each replica $(\kw,i)$ of an unencrypted key $\kw$ is protected by applying the pseudorandom function~$F$, discussed in \S\ref{ssec:prelim}, to the replica identifier to generate an encrypted label $F(\kw,i)$ for the replica. In the rest of the paper, we refer to the original unencrypted key as the plaintext key, and the encrypted label for each replica as the ciphertext key.
To remove the 
remaining non-uniformity, ``fake'' queries are added: these queries are sampled from a carefully 
crafted fake access distribution $\fakedist$ to boost the likelihood of accessing replicated KV pairs, until the resulting distribution is uniform. 
Security requires ensuring that fake and real queries be indistinguishable; to achieve 
this, encrypted queries are issued in small batches of size $B$, where each query is either 
real or fake with equal probability. Since the adversary cannot observe traffic between the clients and the proxy server, it has no way to distinguish real and fake queries within any batch. To prevent an adversary from distinguishing between reads and writes, every access is performed as a read followed by write of a freshly encrypted value. 
Writes to keys with multiple replicas could reveal which replicas belong to the same key; thus, only one replica is updated at the time of the write query, and the write value is cached at the proxy in a data structure called the \updatecache, and the remaining replicas are opportunistically updated during subsequent fake or real queries to the replicas.

Dynamic adaptation to changes in the underlying access distribution is achieved by adjusting the fake-distribution ($\fakedist$), and by reassigning the number of replicas across keys. This can be done securely by exploiting the observation that the total number of replicas 
is exactly $2n$, regardless of the underlying distribution. As such, when the distribution changes, for every key 
that must lose a replica, another must gain a replica to ensure the distribution remains smooth.  
Thus, replicas can be reassigned opportunistically for all such key-pairs
using a replica-swapping protocol. 

In summary, to enable oblivious data access for the general case of read/write workloads and for time-varying distributions, Pancake uses a centralized, {\em stateful}, proxy that stores (1) the \updatecache to buffer writes until they are opportunistically propagated to all the replicas; (2) distribution-related state; and (3) replica-related state, to execute the replica swapping process during distribution changes. Using this state, Pancake enables oblivious data access by performing three tasks at the proxy in failure-free scenarios: (1) generating ``fake'' queries for each real client query; (2) updating \updatecache upon each query; and (3) issuing a batch of queries comprising real and fake queries to the server, and relaying the response for the real query back to the client.

\section{Limitations of Strawman approaches}
\label{sec:strawman}
In this section, we describe subtle security vulnerabilities with strawman approaches to designing distributed, fault-tolerant, systems for oblivious data access. 

\subsection{Centralized proxy: Insecure and/or long periods of unavailability}
\label{ssec:beefy-single-proxy}
The stateful nature of the centralized proxy makes it challenging to simultaneously achieve oblivious data access security guarantees, availability and scalability upon a failure. If achieving scalability were the only goal, the proxy server could be overprovisioned with large bandwidth and/or compute resources; however, achieving security and availability upon a failure is hard due to the proxy being stateful: the na\"ive solution of replacing the failed proxy server with a new one and having clients reissue failed queries results in violating security and correctness guarantees:

\vspace{-0.05in}
\begin{itemize}[itemsep=0pt, leftmargin=*]
  \item Consider the (simplest) case of a read-only workload with a static access distribution. Replacing a failed proxy server with a new one, and having clients reissue the failed queries, results in the following subtle security issue. Consider a real query on key $k$; and consider the scenario where the proxy fails in the middle of sending out queries (both real and fake) in the batch to the KV store, that is, some of the queries in the batch have been sent out while others are lost. Since the proxy has failed, the client would receive no response for $k$; thus, upon restarting the proxy, the client will retry a real query on $k$. The retried queries will result in the same real accesses, but potentially new fake accesses. An adversary can thus exploit the transcript of queries at the server to identify real queries with high confidence by isolating repeated accesses right before and right after a failure, hence gaining sensitive information.
  \item Write queries make the problem significantly more challenging. Consider a write query to a key with two replicas; suppose the proxy fails when the write value has propagated to only one of the replicas (and thus, is buffered in the UpdateCache waiting to be propagated to the other replica). We now replace the failed proxy with a new one. Since the UpdateCache state is lost, when a read query for this key is received, the new proxy could end up reading the value from one of the stale replicas, violating the data correctness/consistency guarantee. Alternatively, if the new proxy reads all replicas of the key to determine which one has the latest value (\eg, using timestamps), oblivious data access guarantees would be violated since an adversary can identify replica correlations (replicas being accessed belong to the same key) by analyzing queries right after a failure. 
\end{itemize}

\noindent
For a centralized stateful proxy design and for the general case of read/write workloads over time-varying distributions, to avoid the above security and correctness violations upon a failure, the proxy state must be reconstructed---\eg, by downloading all the data from the cloud service, decrypting the data, reinitializing the data structures, re-encrypting the new data structures, and writing all the new data back to the server---before issuing new queries. Even for moderate-sized KV stores, this would incur extremely large bandwidth and compute overheads, as well as long unavailability periods. 

In summary, replacing the centralized proxy server with a new one (upon a failure) and having clients reissue the queries either fails to ensure critical system properties (security and/or correctness), or results in large unavailability periods. This motivates the need for a distributed proxy architecture. 

\begin{figure*}
\begin{minipage}[t]{.38\textwidth}
  \centering
  \includegraphics[width=\columnwidth]{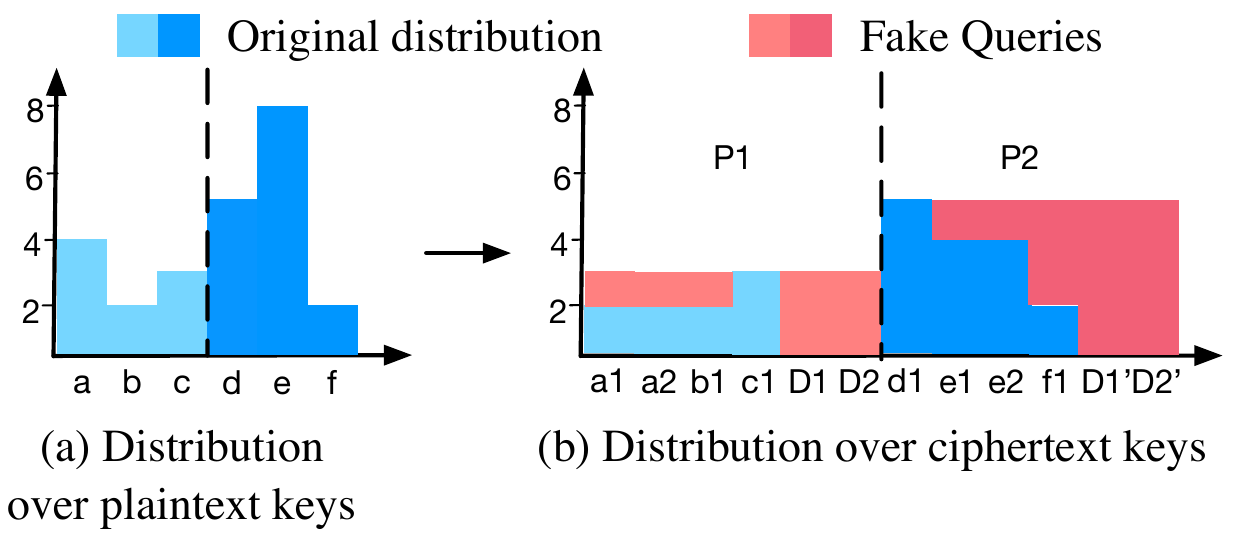}
  \caption{\small Security violation in one-layer approach.
  } 
  \label{fig:dists}
\end{minipage}%
\begin{minipage}[t]{.23\textwidth}
    \centering
    \captionsetup{justification=centering}
    \includegraphics[width=\columnwidth]{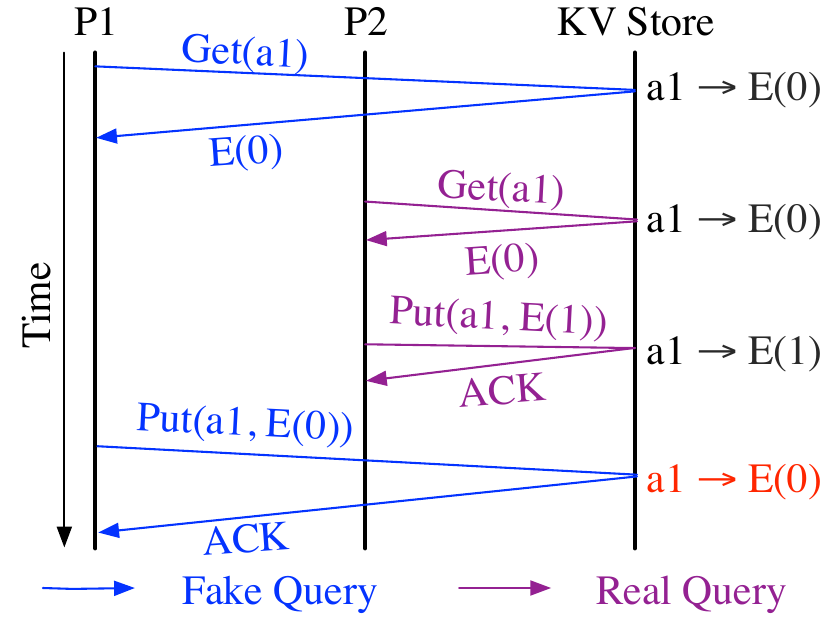}
    \caption{\small Correctness violation in one-layer approach.}
    \label{fig:inconsistency}
    \vspace{-1.5em}
\end{minipage}%
\begin{minipage}[t]{.38\textwidth}
  \centering
  \includegraphics[width=\columnwidth]{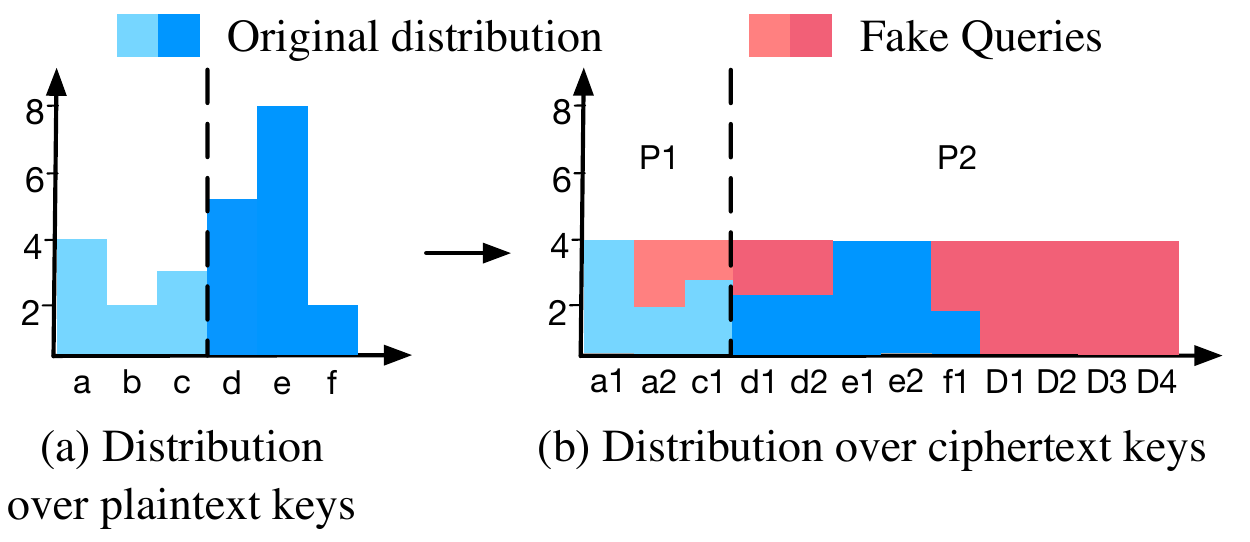}
  \caption{\small Security violation in two-layer approach.
  }\vspace{-0.5em}
  \label{fig:naive2}\vspace{-1em}
\end{minipage}\vspace{-1em}
\end{figure*}

\subsection{Challenges in Distributing Proxy Logic}
\label{ssec:static}
We now describe security and correctness vulnerabilities with na\"ively distributing the proxy state and logic across multiple physical servers. 

\paragraphb{Na\"ively partitioning both the proxy state and the query execution responsibility leads to security violations}
A straightforward approach to designing a distributed proxy for oblivious data access is to partition both the proxy state and the query execution responsibility across multiple physical servers---each proxy server stores the UpdateCache and access distribution for a subset of the plaintext keys (\eg, using hash partitioning over the plaintext keys); clients forward their (real) query on key $k$ to the proxy server responsible for $k$; and, upon receiving a real query, the proxy server generates fake queries based on distribution {\em corresponding to its own partition}, and executes these queries on the storage service.

While this approach scales linearly with number of physical proxy servers, it suffers from security vulnerability. In particular, it does not guarantee that the resulting distribution observed by the adversary is independent of the input distribution. Consider the scenario shown in Figure~\ref{fig:dists}~(a). Here, plaintext keys are partitioned across two proxy servers---P1 is responsible for keys \{$a$, $b$, $c$\}, and P2 is responsible for keys \{$d$, $e$, $f$\}. Since, each proxy server operates only on its local plaintext key partition, P1 selectively replicates key $a$ into 2 replicas $a_{1}$, $a_{2}$, and introduces two dummy key replicas $D_{1}$, $D_{2}$, leading to a total of $6$ ciphertext keys; it then adds fake queries to make the access distribution across these ciphertext keys uniform. Similarly, P2 selectively replicates key $e$ into 2 replicas $e_{1}$, $e_{2}$, and introduces two dummy key replicas $D_{1}'$, $D_{2}'$, again leading to a total of $6$ ciphertext keys; P2 then adds fake queries to make the access distribution across these ciphertext keys uniform. Figure~\ref{fig:dists}~(b) shows the final output access distribution over ciphertext keys. Since P1 and P2 smooth the distribution over their sets of plaintext keys independently, and since the key set assigned to P2 has a higher average access frequency than the key set assigned to P1, the frequency of accesses over ciphertext keys for P2 is higher than the frequency of accesses over ciphertext keys for P1. In particular, the overall access distribution over all ciphertext keys is dependent on the input distribution over the two subset of keys, thus leaking sensitive information. 

\paragraphb{Replicating proxy state across all physical servers but na\"ively partitioning query execution responsibility leads to security violations} To avoid the security vulnerability in the previous scenario, one possible approach is to replicate the entire proxy state (UpdateCache and access distribution) across all physical servers in the distributed proxy. We will need to keep the state consistent across all physical servers---various mechanisms exist for this; for instance, clients can broadcast each query to each physical server to keep the access distribution consistent, and servers could use a distributed protocol (\eg, state machine replication) to keep the UpdateCache consistent. Let us ignore the scalability issues with maintaining such consistent state for a moment.

To avoid bandwidth and compute bottleneck, we still want each query to be executed at one (or a small number) of the physical proxy servers. Thus, each physical server will now be responsible for receiving real queries from the clients for a subset of the keys (again, \eg, using hash partitioning over the plaintext keys), and generating fake queries for each real query (now on the entire distribution). One question remains: which physical server should send the (real and fake) queries to the storage service on the cloud? Unfortunately, both the obvious solutions---the server generating the batch executes all queries in the batch, and the server responsible for plaintext key $k$ executes all (real and fake) queries for the key $k$ (independent of which server generated the fake query)---suffer from security and/or correctness vulnerabilities.

To see the issue with the first solution, consider the example in Figure~\ref{fig:inconsistency} with two proxy servers P1 and P2: to serve a client query to write value $1$ to key $a$, P2 sends a get($F(a, 1)$) followed by put($F(a, 1)$, $E(1)$) query to the KV store, where $(a, 1)$ is one of the ciphertext key, or replica, corresponding to $a$. At the same time, P1, unaware of P2 ongoing write query, sends a fake put query to the same ciphertext key $(a, 1)$ in response to another client query. Based on the timeline 
of operations shown in Figure~\ref{fig:inconsistency}, the fake put from P1 overwrites the real put from P2, resulting in incorrect system behavior. Note that the incorrectness occurs since two different proxy servers issue queries for the 
same ciphertext key.

Unfortunately, the second solution also suffers from security vulnerabilities---partitioning the query execution across physical servers reveals not only which {\em plaintext} keys are managed by each server, but also their relative access frequencies. Figure~\ref{fig:naive2} shows an example; the scenario is the same as Figure~\ref{fig:dists}, but with selective replication and fake query generation done over the entire distribution across all plaintext keys---thus, as shown in Figure~\ref{fig:naive2}~(b), in addition to selective replication of keys $d$ and $e$, $4$ dummy key ($D$) replicas ($D_1$, $D_2$, $D_3$, $D_4$) were added, and the access distribution across ciphertext keys is uniform. We use the same partitioning of plaintext keys across P1 and P2 as in the example of Figure~\ref{fig:dists}---P1 handles all real and fake queries for the three less popular plaintext keys, while P2 handles all queries for the three more popular plaintext keys and the dummy key. The challenge, however, is that although each server handles roughly equal number of plaintext keys, the number of ciphertext keys handled by P1 ($=3$) and P2 ($=9$) are very different. This leaks the subset of keys handled by each server and, by extension, their relative popularities to the adversary. Even if the volume of traffic issued by individual proxy servers is hidden (\eg, via a trusted gateway/NAT so that all proxy servers have the same publicly visible IP address), failures of one of the physical proxy servers would reveal the same information. Even if clients were to retry their queries upon a failure, in-flight queries to the KV store from a failed server would be repeated, again revealing the
same information. 

\paragraphb{Summary} The above discussion leads to three different design principles for distributed, fault-tolerant, oblivious data access systems. From the partitioning-based approach, we learn that, to achieve oblivious data access, each physical server in the distributed proxy should perform selective replication and (fake) query generation over the entire distribution across all plaintext keys (thus, each physical server should know the access distribution across the entire set of plaintext keys). The replication-based approach leads to two additional principles. First, even if proxy state can be replicated in a consistent and scalable manner, maintaining correctness requires that no two physical proxy servers should send the queries for the same ciphertext key; in other words, query execution should be partitioned by ciphertext keys across different physical servers. Second, to avoid security vulnerability, no single proxy server should be deterministically responsible for executing queries for all ciphertext keys corresponding to the same plaintext key; that is, query execution should be partitioned by ciphertext keys---randomly, and independent of plaintext keys---across physical proxy servers.

\section{\name Design}
\label{fastpancake}
\label{sec:design}
We now present the \name distributed, fault-tolerant, proxy architecture. 

\subsection{Design Overview}
\label{ssec:design-overview}

\begin{figure}[t]
    \centering
    \includegraphics[width=\linewidth]{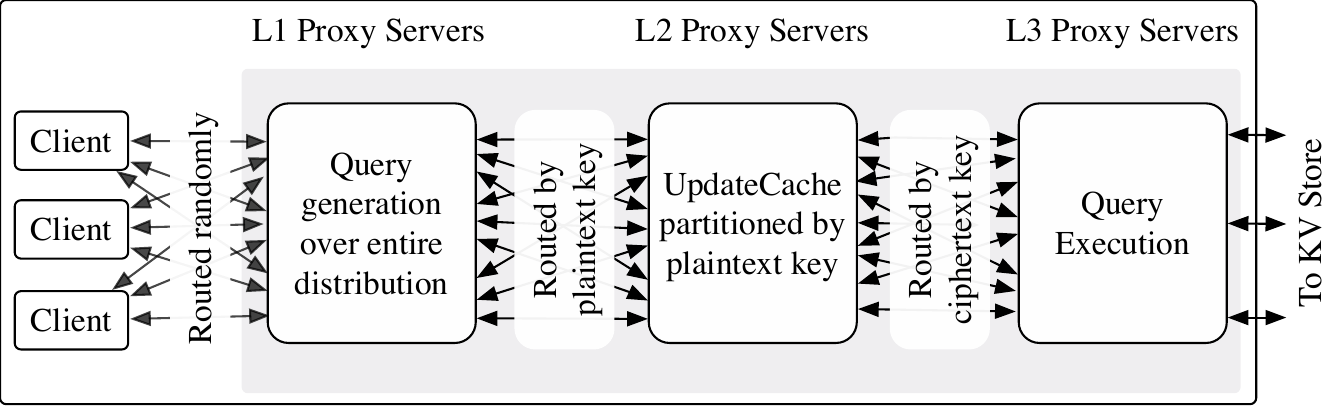}
    \caption{An overview of three-layer \name architecture
    }
    \label{fig:threelayer}\vspace{-1em}
\end{figure}

\name uses a novel layered architecture, with three {\em logical} layers{\footnote{On a lighter note, our work seems to formally establish the widely-agreed belief that three is the right number for a \name~\cite{shortstack-ihop}!}}, as shown in Figure~\ref{fig:threelayer}. Each layer has multiple logical proxy servers for fault tolerance and/or scalability purposes, and embodies one of the three design principles outlined at the end of the previous subsection. In the first layer (\li), proxy servers are responsible for a random subset of client queries---upon receiving a real client query on a plaintext key, the server generates real and fake queries (over ciphertext keys); importantly, fake queries are generated using the {\em entire} access distribution across all plaintext keys. In the second layer, \lii, proxy servers are responsible for maintaining a partition of the \updatecache state; importantly, the \updatecache is partitioned by {\em plaintext} keys across the \lii servers. Finally, in the third layer, \liii, each proxy server is responsible to execute real and fake queries on the KV store for a random, distinct, subset of {\em ciphertext} keys.

We outline the lifetime of a query with the layered \name architecture in a {\em failure-free scenario}. The client sends the query to a randomly selected \li proxy server; the \li server generates the batch comprising real and fake queries (recall, these generated queries are on ciphertext keys). The \li server then forwards each individual query within the batch to one of the \lii servers---the one that maintains \updatecache state for the corresponding plaintext key in the query. Upon receiving a query, an \lii server updates its local partition of the \updatecache, and forwards the query to one of the \liii servers---the one that is responsible for executing queries for that ciphertext key. The \liii server ultimately forwards the query to the KV store; upon receiving a response from the KV store, the \liii server sends a response for the real query back to the client, as well as an acknowledgement in the reverse direction of the original path taken by the query (from \liii to \lii to \li) to clear any buffered state associated with the query. We provide more details on the three-layer \name architecture and query execution in \S\ref{ssec:three-layer}.

For fault-tolerance against $f$ failures, each of the \li and \lii proxy servers use $f+1$ replication along with the chain replication protocol~\cite{chain-replication}. Replicating \li servers prevents the security vulnerabilities discussed in \S\ref{ssec:beefy-single-proxy} that are caused by clients retrying queries upon failures. Specifically, as we discuss in \S\ref{ssec:failures}, \name uses chain replication to guarantee that a batch of queries is never partially executed---either all the queries in a batch are (eventually) forwarded to the KV store or none of them are---thus preventing access pattern leakage even when failures happen. Replicating \lii servers prevents \updatecache state from being lost due to failures. As we will show, \liii server failures do not have the same security vulnerability as \li and \lii server failures. Thus, \liii layer is not chain replicated; however, it needs at least $f+1$ servers to maintain availability during failures---if one of the \liii server fails, the remaining \liii servers take over its load. Upon an \liii server failure, in-flight queries at the server will be dropped and \lii servers will reissue the dropped queries. While this results in duplicate queries being forwarded to the KV store, we will show that these duplicate queries do not reveal any sensitive information---the adversary would only observe duplication of queries to a random subset of labels independent of the input distribution. We provide more details on \name mechanisms for handling failures in \S\ref{ssec:failures}.

\begin{figure}
  \centering
  \includegraphics[width=\columnwidth]{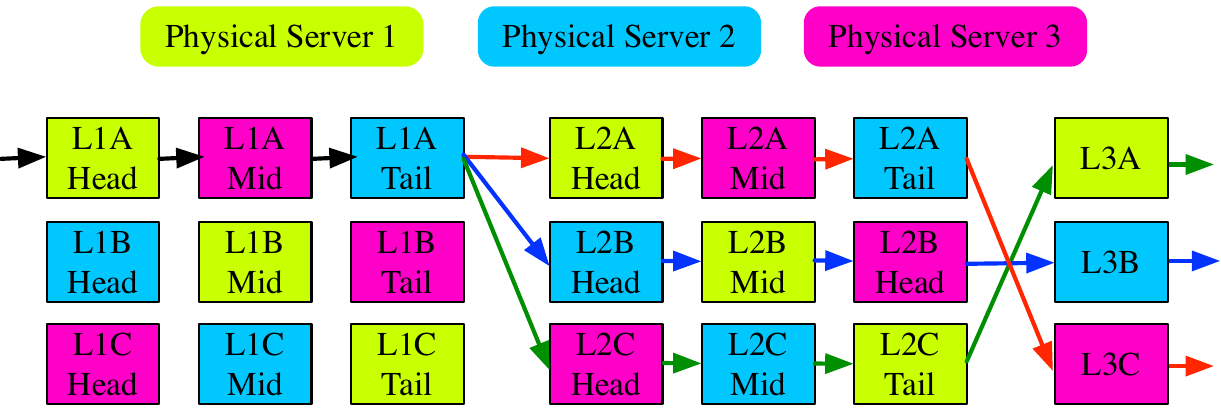}
  \caption{An instantiation of \name's three-layer architecture that guarantees system security and availability with $f=2$ failures, and achieves $k=3\times$ scalability (as defined in \S\ref{ssec:design-overview}). Multiple logical layers in \name are colocated on the same physical server. The arrows depict the lifetime of a single query.}
  \label{fig:chain_replication}\vspace{-1.5em}
\end{figure}

\name design allows independently setting desired fault tolerance $f$ and scalability factor $k$. Specifically, to achieve a factor $k$ scalability---that is, achieving system throughput a factor of $k$ higher than a centralized proxy---along with fault tolerance against $f$ failures, \name creates $k$ independent \li and \lii chains that operate in parallel. The case of \liii is again different; if $f+1 > k$, \name will already have at least $k$ \liii proxy servers to ensure fault tolerance (as described earlier). For $f+1 \leq k$, \name uses a total of $k$ \liii proxy servers, thus guaranteeing both fault tolerance against $f$ failures and a factor $k$ scalability. Figure~\ref{fig:chain_replication} shows an example for $f=2$ and $k=3$. 

\paragraphb{Colocating \name logical layers and their replicas on a small number of physical servers} Consider an instantiation of \name with fault-tolerance against up to $f=2$ failures and $k=3$ scalability. Then, given the above design, \name will require $3$ \li and $3$ \lii chains, each having $3$ logical replicas within the chain replication protocol. In addition, \name will require $3$ logical servers in the \liii layer. Overall, for $f=2$ and $k=3$, \name requires $21$ ``logical'' units. However, as shown in Figure~\ref{fig:chain_replication}, all these $21$ logical units can be packed on $3$ physical servers without compromising security, fault tolerance, availability and scalability. In particular, the replicas of each logical server in each layer are staggered across the physical servers such that no two replicas of the same logical server within the same layer are co-located on the same physical server. Hence, even upon failure of any two physical servers, one replica from each of the \li servers, one replica from each of the \lii servers, and one \liii server will still be alive, ensuring security, availability and $f=2$ fault tolerance. In general, using a technique from~\cite{hibari}, \name achieves a factor $k$ scalability and fault tolerance against $f \leq k-1$ failures, using only $k$ physical servers. Since any system that tolerates $f$ failures and achieves a factor $k$ scalability must require at least $\max(f+1, k)$ physical servers, \name uses minimum resources to provide these properties.

We provide more details on design of each layer in the \name layered architecture, the mechanisms for fault tolerance, and the mechanisms for handling dynamic distributions in \S\ref{ssec:three-layer}, \S\ref{ssec:failures} and \S\ref{ssec:dynamic_dist}, respectively. 

\begin{figure*}[t]
  \centering
  {\setlength{\fboxsep}{.1\fboxsep}
    \framebox{
      \begin{tabular}{llll}
      \begin{minipage}[t]{.21\textwidth}\gamesfontsize\setstretch{1.1}
        \procedurev{Init$(\estdist,\DB, S, f)$}\\
        $\DB', \fakedist \gets \ninj.\Init(\DB, \estdist)$\\
        $S_{\li}, S_{\lii}, S_{\liii} \gets \Configure(S)$\\
        $\delta \gets \Weights(S_{\lii}, \DB')$\\
        \textbf{return} $\DB', (S_{\li}, S_{\lii}, S_{\liii}), \delta$\\
      \end{minipage}
      &
      \begin{minipage}[t]{.175\textwidth}\gamesfontsize\setstretch{1.1}
        \procedurev{$s_{\li}$.ProcessQuery$(\kw,\val)$}\\
        $\ell \gets \ninj.\Batch(k)$\\
        For $((k, j), \val) \in \ell$:\\
        \ind $s_{\lii} \gets S_{\lii}[\mathcal{H}(k)]$\\
        \ind $s_{\lii}$.Enqueue($(k, j, \val)$)\\
      \end{minipage}
      &
      \begin{minipage}[t]{.19\textwidth}\gamesfontsize\setstretch{1.1}
        \procedurev{$s_{\lii}$.Process()}\\
        $\kw, j, \val \gets$ Dequeue()\\
        $\val \gets$ $\ninj.\updatecache(\kw, j, \val)$\\
        $s_{\liii} \gets S_{\liii}[\mathcal{H}(F(\kw, j))]$\\
        $s_{\liii}$.Enqueue($s_{\lii}$, $(F(\kw, j), \val)$)\\
     \end{minipage}
     &
     \begin{minipage}[t]{.205\textwidth}\gamesfontsize\setstretch{1.1}
        \procedurev{$s_{\liii}$.Process($\delta$)}\\
        $s_{\lii} \getsbiased{\delta} S_{\lii}$\\
        $\kw', \val \gets$ Dequeue($s_{\lii}$)\\
        $\val \gets $ ReadThenWrite($\DB'$, $\kw'$, $\val$)\\
        \textbf{return} $k',\val$\\
     \end{minipage}
    \end{tabular}
    }
  }\vspace*{-.25em}
  
  \caption{\small\textbf{\name initialization and processing logic at \li, \lii and \liii servers.} $S_{\li}, S_{\lii}, S_{\liii}$ are the sets of proxy servers in each layer, and $S$ is the set of physical servers upon which they are initialized. ($k$, $v$) corresponds to the plaintext key-value pair, while $j$ is the replica identifier for a given replica of the key. $F$ is a secretly keyed pseudorandom function and $\mathcal{H}$ is a consistent hash function.
  }
  \label{fig:init_logic}
  \label{fig:l1_logic}
  \label{fig:l2_logic}
  \label{fig:l3_logic}
  \label{fig:static_algo}
  
  \vspace{-1.5em}
\end{figure*}

\begin{figure}
  \centering
  \includegraphics[width=\linewidth]{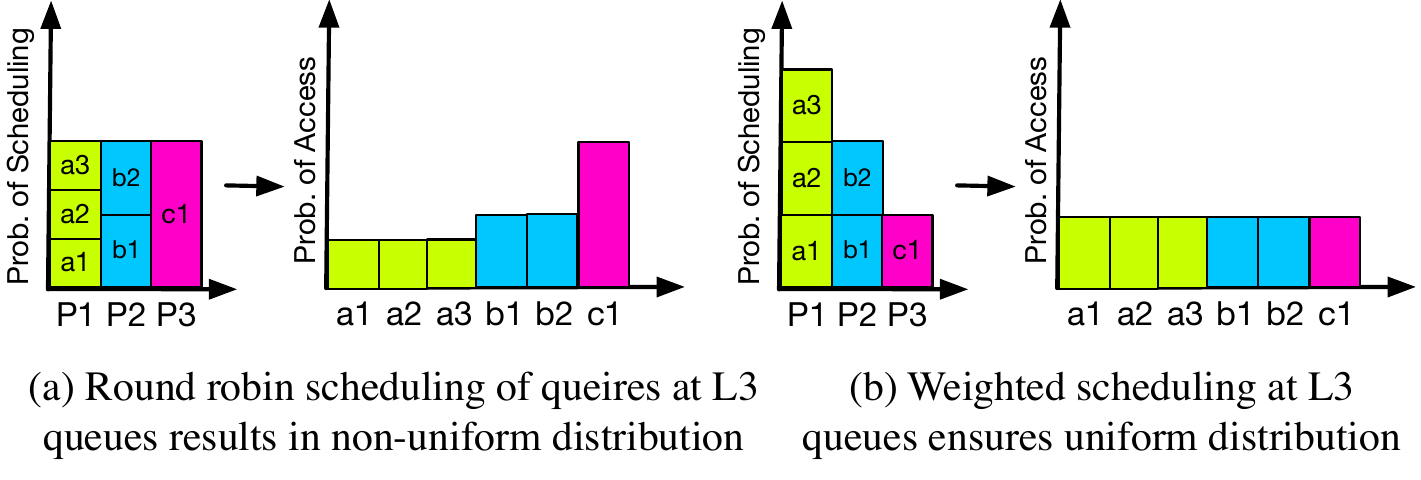}
  \caption{Query-scheduling at \liii layer should ensure uniform distribution over ciphertext keys for security. In each figure, (left) shows the probability of scheduling queries from each of the \lii servers, while (right) shows the resulting distribution across ciphertext keys.
  }
  \label{fig:request-scheduling}\vspace{-1.25em}
\end{figure}

\subsection{\name Design Details}
\label{ssec:three-layer}
In this subsection, we describe \name's three-layer architecture in detail, for the case of no failures and static access distribution. We will extend this design to handle failures and dynamic distributions in \S\ref{sec:failures} and \S\ref{ssec:dynamic_dist}, respectively.

In a failure-free scenario, the key challenge that \name addresses relative to a single proxy architecture is scalability. To achieve $k$-factor scalability in the failure-free scenario, \name uses $k$ (logical) proxy servers in each layer. For example, in Figure~\ref{fig:chain_replication}, each layer would consist of three nodes, \eg, L1A, L1B and L1C for the \li layer, L2A, L2B and L2C for the \lii layer, and L3A, L3B and L3C for the \liii layer.

\paragraphb{Details of three-layer operation} Figure~\ref{fig:l1_logic} details the precise initialization and \li/\lii/\liii server logic in \name. \name employs the following functionalities from \pancake ($\ninj$)~\cite{pancake} as a black-box:
\vspace{-0.1in}
\begin{itemize}[itemsep=0pt, leftmargin=*]
  \item an $\Init$ function, which takes as input an estimate of the access distribution $\estdist$ and the unencrypted KV store $\DB$ of size $n$ plaintext keys, and generates an encrypted KV store $\DB'$ of size $2n$ ciphertext keys, along with a fake distribution $\fakedist$ over $\DB'$;
  \item a $\Batch$ function, which takes a query on a plaintext key $k$ in $\DB$ as input, and generates (using $\estdist$ and $\fakedist$) a batch of $B$ ($B=3$ by default) ciphertext queries to $\DB'$; and,
  \item an \updatecache function that internally updates per-plaintext key state, and returns an encrypted (possibly updated) value to be written to the KV store. 
\end{itemize}
\vspace{-0.05in}
\noindent
We now outline how \name distributes the execution of \pancake across its three-layer design:

\paragraphc{Initialization ($\Init()$ in Figure~\ref{fig:l1_logic}):}
\name first performs \pancake initialization (using $\ninj.\Init$), transforming the unencrypted KV store $\DB$ with $n$ plaintext keys to the encrypted KV store $\DB'$ using $2n$ ciphertext keys, using an estimate of the underlying access distribution $\estdist$. During the process, the adversary just observes insert operations of $2n$ ciphertext keys, which does not reveal any information. \name then initializes and configures $k$ logical proxy servers in each of the three layers on top of $k$ physical servers. Finally, \name computes a weight vector $\delta$, containing weights assigned to each \lii server (proportional to the volume of ciphertext traffic generated by it). As will be discussed, \liii servers use these weights to process \lii queries such that the queries issued by \liii servers appear uniform random (recall, this subsection focuses on failure-free scenario, where \name achieves uniform random distribution over ciphertext keys).

\paragraphc{Query processing logic ($s_{\li}.\ProcessRequest()$, $s_{\lii}.\Process()$ and $s_{\liii}.\Process()$ in Figure~\ref{fig:l1_logic}):} Clients forward each query to a randomly chosen \li proxy server. Upon receiving a query, the \li server generates a batch of $B$ queries (using $\ninj.\Batch$) that comprises both real and fake queries to $\DB'$. The \li server then enqueues each query in the batch across different \lii servers based on the hash of the query's plaintext key. 

Upon receiving a query, an \lii server calls $\ninj.\updatecache$ which leads to two sequential actions. First, the per-plaintext key state stored at the \lii server is updated; and second, if this query can be used to propagate outstanding write queries into the plaintext key replicas, the value to be written to the KV store is updated. It then forwards the query to the corresponding \liii server based on the hash of the query's ciphertext key (denoted as ``Enqueue'' in Figure~\ref{fig:l1_logic}).

Finally, each \liii server maintains a separate queue for each \lii server it receives queries from, and dequeues queries from the queues following a biased distribution determined by the weight vector $\delta$. To hide whether the query is a read or a write, \name employs the standard approach from prior oblivious data access schemes of performing each queries as a read followed by a write to the KV store. Specifically, in Figure~\ref{fig:static_algo}'s ReadThenWrite() method, the \liii server first reads and decrypts the value associated with the query from the key-value store. If the value needs to be updated (\ie, write query), then the plaintext value is updated accordingly. Finally, the \liii server writes the (re)encrypted value for the query back to the KV store.

\paragraphb{Query scheduling at \liii layer for security} The way in which queries from different \lii servers are scheduled at each \liii server has security implications. As a concrete example, consider a scenario where three plaintext keys $a$, $b$ and $c$ with $6$, $4$ and $2$ replicas (or, ciphertext keys), respectively, are mapped to three different \lii servers $P_1$, $P_2$ and $P_3$. Suppose we have two \liii servers, and one of these handles half of the ciphertext keys for each plaintext key (Figure~\ref{fig:request-scheduling} illustrates the example, focusing only on one of the \liii servers and the ciphertext keys mapped to this server).
If the \liii server processes queries from each server with equal likelihood (\eg, using round-robin scheduling), then the distribution across ciphertext keys would no longer be uniform, since queries from the first server would be under-sampled while those from the third server would be over-sampled (Figure~\ref{fig:request-scheduling}~(a)). To ensure \liii servers still issue queries that are uniform random, they maintain a separate query queue for each \lii server, and process the queues in proportion to the volume of traffic the corresponding \lii servers generate. In the above example, the \liii server would schedule queries from each of the \lii servers with probabilities $3/6$, $2/6$ and $1/6$, respectively, leading to a uniform distribution across ciphertext keys (Figure~\ref{fig:request-scheduling}~(b)).

\paragraphb{Accurately estimating the access distribution}
\name employs a lightweight mechanism through which a single \li proxy server can observe all client queries, enabling distribution estimation as accurately as a centralized proxy system~\cite{pancake}.
One of the \li proxy servers, designated as the leader, monitors the access distribution (handling failures will be discussed in the next subsection). Upon receiving a query, an \li proxy server asynchronously forwards the corresponding plaintext key---and not the entire query---to the \li leader, ensuring that the leader has a complete view of the access distribution. Sending the plaintext key and not the entire query to the leader is an useful optimization for both read and write queries---it reduces the additional network load (for write queries, values are not forwarded; for read queries, the responses are not forwarded) since the plaintext key is typically much smaller than the value itself (\eg, $8$B keys for $1$KB value in~\cite{ycsb}). As such, this has negligible impact on \name scalability and performance.

\subsection{Handling Failures}
\label{sec:failures}
\label{ssec:failures}

We now describe how \name ensures fault-tolerance while preserving security and correctness under failures. We assume the standard \textit{fail-stop} failure model~\cite{fail-stop, chain-replication}.
\name employs a separate centralized \textit{coordinator} node which keeps track of the health of the proxy servers using heartbeats, detects failures, and notifies other proxy servers as needed to designate a fail-over node. The coordinator node is also replicated using ZooKeeper~\cite{zookeeper} for strong consistency.
As such, a $(2r+1)$-replicated coordinator can tolerate up to $r$ failures without any security or performance consequences. 

\paragraphb{Handling \li and \lii failures} Failure of a single \li server does not impact the availability of \name, as future client queries could potentially be load balanced across the remaining \li servers. Such a failure, however, has security implications---consider the case where an \li proxy server fails in the middle of forwarding a batch of queries, \ie, some of the queries in the batch have been forwarded, but others are lost due to the failure. Any real queries that are lost would need to be retried by clients. The retried queries would now result in the same real accesses, but with \textit{new} fake accesses generated. This permits an adversary to identify real queries with high confidence by simply isolating the repeated accesses due to failures. To protect against such a vulnerability, \name ensures the following invariant:
\begin{invariant}[Batch atomicity]\label{inv:batch_atomicity}
  Either all of the queries in a batch are forwarded to the KV store, or none of them are.
\end{invariant}

\noindent
\name achieves this by replicating the state of the \li proxy servers across multiple replicas ($f+1$ replicas to tolerate up to $f$ failures) using chain replication protocol~\cite{chain-replication}.
As shown in Figure~\ref{fig:chain_replication}, \name maintains staggered chains across a fixed pool of physical servers, such that each physical server hosts the head node of a single chain. Chain replication ensures that all \li replicas in the chain buffer a batch of queries, before the queries are forwarded to \lii servers---the buffered batches are only cleared when all corresponding acknowledgements are received from the \lii servers. As such, as long as any \li replica in the chain is online, the set of buffered batches is available and can be used to retry queries as required, ensuring \cref{inv:batch_atomicity}. 

Since \lii servers store \updatecache partitions, ensuring fault-tolerance for them is crucial for availability, correctness and security. As such, \name replicates the \updatecache state for any key across multiple \lii proxy replicas using chain replication, similar to \li servers.

Within each chain of \li and \lii servers, failures of replicas are handled as per the standard chain replication protocol~\cite{chain-replication}. Since the \li server chains interact with the \lii server chains, additional failure handling is necessary in certain cases. Consider the interaction between an \li tail and an \lii head: if the \li tail fails, its predecessor in the chain becomes the new tail, and resends the queries in the buffered (unacknowledged) batches to the corresponding \lii head replicas. The \lii servers, on the other hand, discard the queries that they have already seen and forward the remaining down the chain. \name facilitates the detection of duplicate queries by assigning unique sequence numbers to each query. If an \lii head fails, on the other hand, its successor become the new head. All \li tails then examine their buffered batches to resend queries that were destined to the failed \lii head. As before, the new \lii head simply discards any queries that it has already seen, forwarding the remaining down the chain.

\paragraphb{Handling \liii failures}
Unlike \li and \lii servers, \liii servers are not replicated, and hence entail different failure handling. Since \liii servers are stateless, if an \liii server fails, the remaining \liii servers can assume the responsibility of the ciphertext labels that the failed server was handling. Since the system remains available as long as at least one of the \liii servers is online, we need at least $f+1$ \liii servers to tolerate $f$ failures. However, there are two subtle issues that can arise due to \liii failures---we describe these next, along with how \name addresses them.

On an \liii server failure, queries that were in-flight at the failed \liii server would be lost, which can then be retried by the \lii servers. Note that such retries can cause duplicate queries being sent to the KV store. Since the duplicate queries are to uniformly accessed ciphertext keys, it may seem like they do not reveal any distribution-sensitive information. However, repeating the queries in exactly the same order (or a correlated order) introduces a subtle security vulnerability. Specifically, when an \liii server fails, \lii tail servers repeat buffered queries (which are uniform random) and redistribute them to different \liii servers. If the order of these queries is exactly the same as before, an adversary can identify the sequences of repeated queries and correlate them to the \lii server that generated those queries. Moreover, the adversary can also map the specific ciphertext keys corresponding to the plaintext keys managed by a particular \lii server, revealing distribution sensitive information. To prevent this leakage, \name \textit{randomly shuffles} buffered queries before repeating them---we formally prove in\iffullversion  ~Appendix~\ref{app:proofs}\else~\cite{suppl}\fi~ how this ensures security under \liii server failures. 

Recall that the \liii server performs a read followed by a write for all queries. For read queries (fake or real), the write simply writes back the value read from the KV store, \ie, a \textit{fake} write. This can lead to consistency issues during failure of \liii servers---fake in-flight write queries sent by a failed \liii server prior to failure could be delayed by the network and overwrite a real write query sent by the new \liii proxy server responsible for the same ciphertext key. To address this issue, after an \liii failure, the \lii servers delay repeating buffered queries for a fixed amount of time to allow potential in-flight queries from the failed \liii server to get delivered to the KV store. We select the wait time at \lii servers long enough to ensure \textit{all} in-flight queries are propagated to the KV store.


\subsection{Handling Dynamic Distributions}
\label{ssec:dynamic_dist}

Designing distributed, fault-tolerant, oblivious data access systems is challenging when underlying distribution can change over time. We outline two reasons. First, the centralized proxy design (\S\ref{ssec:noise}) relies on having a complete view of the underlying distribution to detect and to react to distribution changes. Detecting the change when queries are spread across multiple proxy servers, and informing other proxy servers about the same, introduces the first challenge. Second, if different proxy servers independently 
initiate and terminate the replica swapping phase at different times, the resulting distribution may not
appear uniform random to an adversary. As such, the adversary may be able to leverage this information to 
identify the keys that may have changed in popularity. We next discuss how \name resolves these challenges.

To detect distribution changes, \name leverages the \li leader, which has visibility of all client queries (\S\ref{ssec:three-layer}). The \li leader is responsible for monitoring the access distribution and employs standard statistical tests to check if there is a change in distribution (\ie, from $\estdist$ to $\newestdist$) similar to \pancake.
Upon detecting a change in distribution, the \li leader initiates the distribution change process.
To ensure security and correctness during distribution change, the \li leader employs a specialized protocol inspired by two-phase commit (2PC)~\cite{2pl} to
facilitate an \textit{atomic} transition from $\estdist$ to $\newestdist$ across all servers in its three-layer
design, both during the initiation and termination of the replica-swapping phase employed by \pancake. Our 2PC-based approach guarantees:
\begin{invariant}[Distribution change atomicity]\label{inv:distchange_atomicity}
    Once any \liii proxy server issues a query according to $\newestdist$, all subsequent queries issued by any \liii server must be according to $\newestdist$.
\end{invariant}
\noindent
In other words, there is an instant of time $t_{c}$ in the protocol's execution, such that: (1) before $t_{c}$, all queries are processed according to the distribution $\estdist$, and (2) after $t_{c}$, all queries are processed using $\newestdist$. This allows us to ensure security for \name even under dynamic distributions, as we detail in \S\ref{sec:security}.
The invariant also ensures consistency during distribution change. In particular, since the change of distribution can result in a change in number of replicas for various plaintext keys, the invariant ensures queries from old and new distributions are not mixed together; this guarantees consistency by ensuring stale replicas from the old distribution are not updated incorrectly due to the new distribution by different \lii proxy servers.
We show that our protocol guarantees the above invariant, with a precise specification in\iffullversion  ~Appendix~\ref{app:2pc}\else~\cite{suppl}\fi. 
Failures during the above protocol are handled transparently by chain replication as \li, \lii servers are chain replicated. This ensures that even with failures during protocol execution, Invariant~\ref{inv:distchange_atomicity} is still preserved. As demonstrated in \S\ref{sec:eval}, \name can recover from such failures quickly enough so as to ensure that their effects are not perceptible to an adversary.


\section{Security Analysis}
\label{sec:security}
This section presents a security model for access pattern attacks on a system with distributed, \textit{fault-tolerant} proxy servers, and a proof that \name achieves security under this model.

\subsection{Need for New Security Definitions}
\label{ssec:secneed}
\label{ssec:discussion}

State-of-the-art \ROR (real-or-random indistinguishability) based security definitions for access pattern attacks~\cite{pancake} are unable to capture the security implications of our distributed proxy setting due to two main reasons. First, \ROR-based definitions focus on indistinguishability between a real and a uniform random distribution (over the entire support). However, as discussed in \S\ref{ssec:static}, we do not yet know whether it is possible to guarantee uniform random distribution over the entire support during failures for \textit{any} distributed proxy architecture. Our \IND-based security model and definitions capture the powerful intuition that uniform random distribution is not even necessary: even though the distribution is non-uniform under failures, the adversary does not gain any \textit{usable} advantage as long as the final distribution is independent of the real distribution. More precisely, our \IND-based security focuses on demonstrating indistinguishability between two arbitrary input distributions. As we will show, under our model, the only information revealed to the adversary is that a failure occurred, information the adversary already possess; it cannot, however, use this information in inferring any information about the underlying distribution itself. While it is not uncommon for \IND security to reduce to \ROR security in many settings, this is clearly not the case in our setting if (and, as we note later, only if) there are failures.

The second reason for needing new security model and definitions is that \ROR-based definitions fail to capture the impact of query reordering on the transcripts observed by an adversary due to (i) distributed query processing, and (ii) worst-case timings of proxy failures. Specifically, a key challenge in demonstrating security lies in precisely capturing the effect of the distributed and failure-prone execution of any scheme in a sequential game-based proof, which the \ROR-based approach omits. We thus have to develop accurate \textit{simulators} that transform distributed query processing to an equivalent sequential one. 
Our model and definitions are not specific to \name, and can be used as templates for any distributed, fault-tolerant, proxy design. 

When there are no failures, our security definition captures the same security guarantees as prior work~\cite{pancake} --- our extensions to the model are required to capture the effect of failures in the distributed proxy setting. In incorporating these extensions, we have only strengthened the adversary.

\subsection{Security Definitions and Proof of Security}

We call our security definition Indistinguishability under Chosen Distribution and Failure Attack, or $\INDCDA$ (Figure~\ref{fig:indcda}). The game $\INDCDA$ is parameterized by bit $b$ (to pick one out of the two given distributions), number of queries $q$, the set of proxy servers $S$ on which the distributed oblivious data access protocol runs, the maximum number of server failures $f$ allowed (similar to classical distributed systems literature that provides fault tolerance up to a fixed number of failures), and two distributions (and their estimates) that the adversary tries to distinguish between.

\begin{figure}[t]
  \centering
    \fpage{.25}{\scriptsize
    \procedurev{$\INDCDA^{\adv}_{b,q,S,f,\dist_0,\hat\dist_0,\dist_1,\hat\dist_1}$} \\
    $\DB, \failures, \stadv  \sample \adv_1(f, S)$ \\
    $(\DB',\chains,\delta) \sample \Init(\hat{\dist_b}, \DB, S, f)$ \\
    For $i$ in $1$ to $q$: \\
    \myind $w \sample \dist_b$ \\
    \myind $W \gets W \cup \{w\}$ \\
    $\tau_{1}, \tau_{2}, ... \gets \Process(W,\chains,\failures,\DB',\delta)$ \\
    $b' \sample \adv_3(\stadv, \DB', \tau_{1}, \tau_{2}, ...)$ \\
    \textbf{return} $b'$
    }
  \caption{\textbf{\INDCDA~security game.}}\label{fig:games}\label{fig:indcda}\vspace{-1.5em}
\end{figure}

The adversary first outputs KV pairs $\DB$ and a queue $\failures$ of at most $f$ failure events. Each failure event $e$ is characterized by the tuple $(n, t, \gamma, r)$, where $n$ is the server in $S$ that fails, $t$ is the time at which the last query is issued by $n$ before failure, $t-\gamma$ is the time at which the last query was acknowledged at $n$ before failure, and $r$ is the failure recovery time. 
Next, the distributed proxy scheme's $\Init$ function generates transformed KV pairs $\DB'$, a set of (potentially replicated) servers $\chains$, and internal state $\delta$ specific to the scheme. For instance, in \name, $\chains$ consists of two sets of replicated server chains (with replication factor $f+1$) corresponding to $\li$ and $\lii$ layers, and a set of $>f$ unreplicated servers for the $\liii$ layer. The state $\delta$ corresponds to weights for $\liii$ servers used in query scheduling, as outlined in \S\ref{ssec:three-layer}. 

After initialization, $q$ queries are drawn from the distribution $\dist_b$ and populated into the vector $W$. The proxy scheme's $\Process$ function takes $W$, $\chains$, $\failures$, $\DB'$ and $\delta$ as input, and generates the output transcripts $\tau$, which is fed to the adversary to try and guess the underlying distribution (\ie, the bit $b$). The adversary ``wins'' if it guesses $b$ correctly. Intuitively, the security goal captured by the definition rules out access pattern attacks since the probability of accessing an encrypted label in $\DB'$ is independent of the underlying distribution itself, and an adversary cannot determine which distribution was used to generate accesses to $\DB'$.

Note that, $\INDCDA$ definition is independent of \name's design. Specifically, our definitions only assume the presence of multiple failure-prone proxy servers which are initialized using an $\Init$ function and process queries using a $\Process$ function, neither of which are specific to \name. Thus, our security model and definitions can be used to study oblivious data access properties of any distributed system that can factor its initialization and query processing logic along these two functions.

\begin{figure*}
  \centering
  \includegraphics[width=0.9\linewidth]{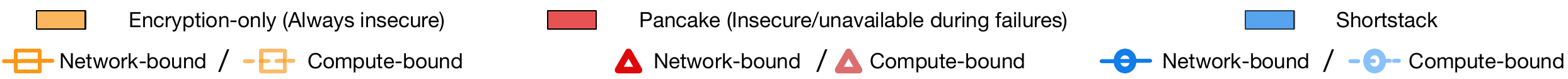}
  \includegraphics[width=0.32\linewidth]{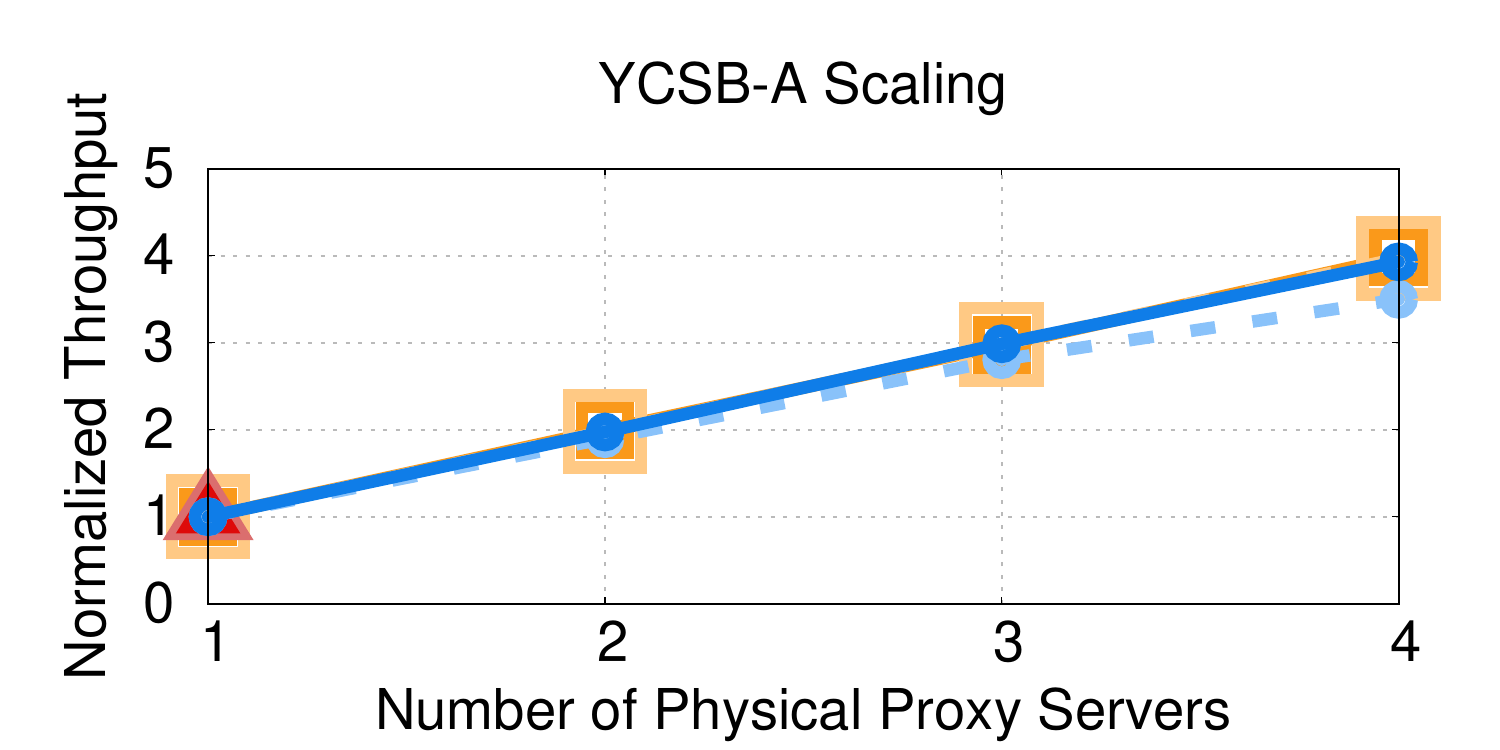}
  \includegraphics[width=0.32\linewidth]{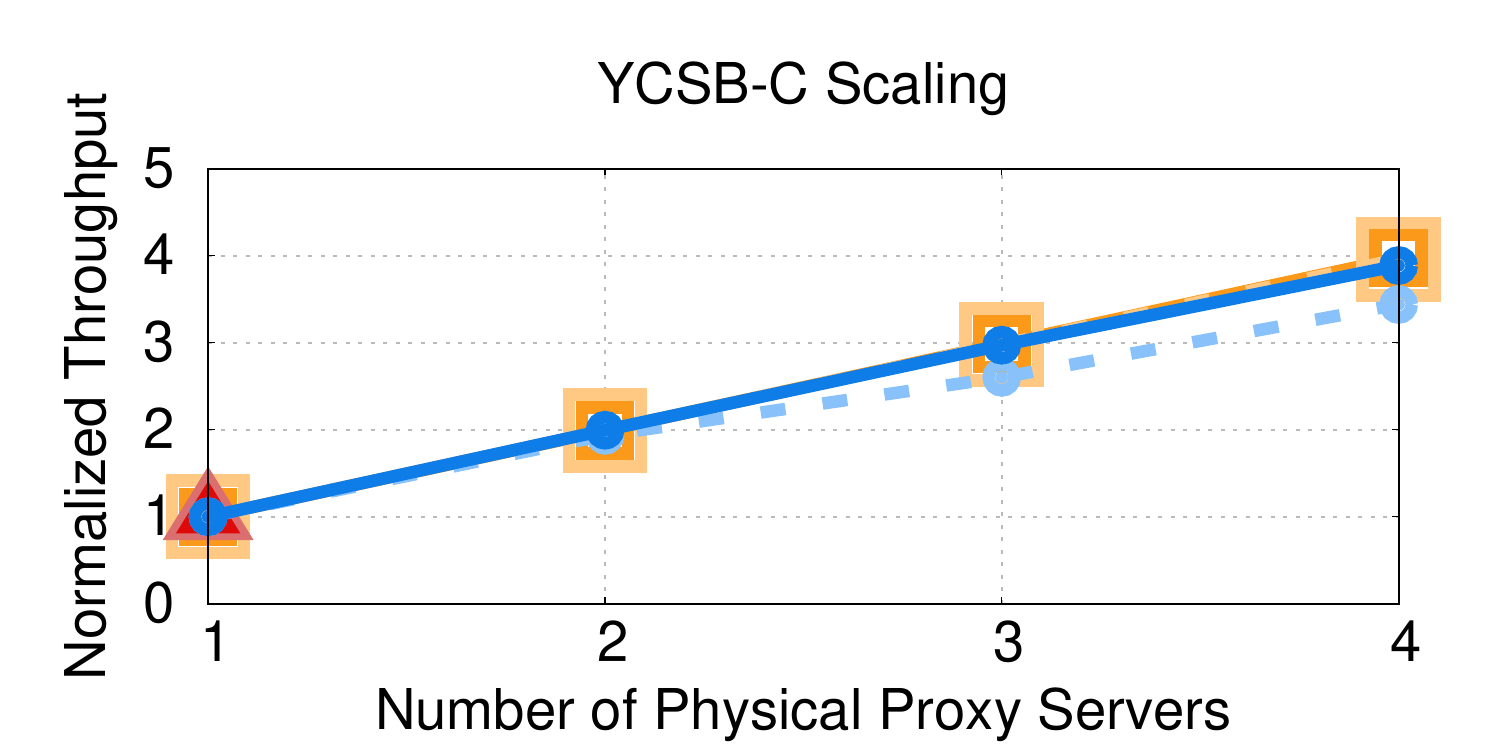}
  \includegraphics[width=0.32\linewidth]{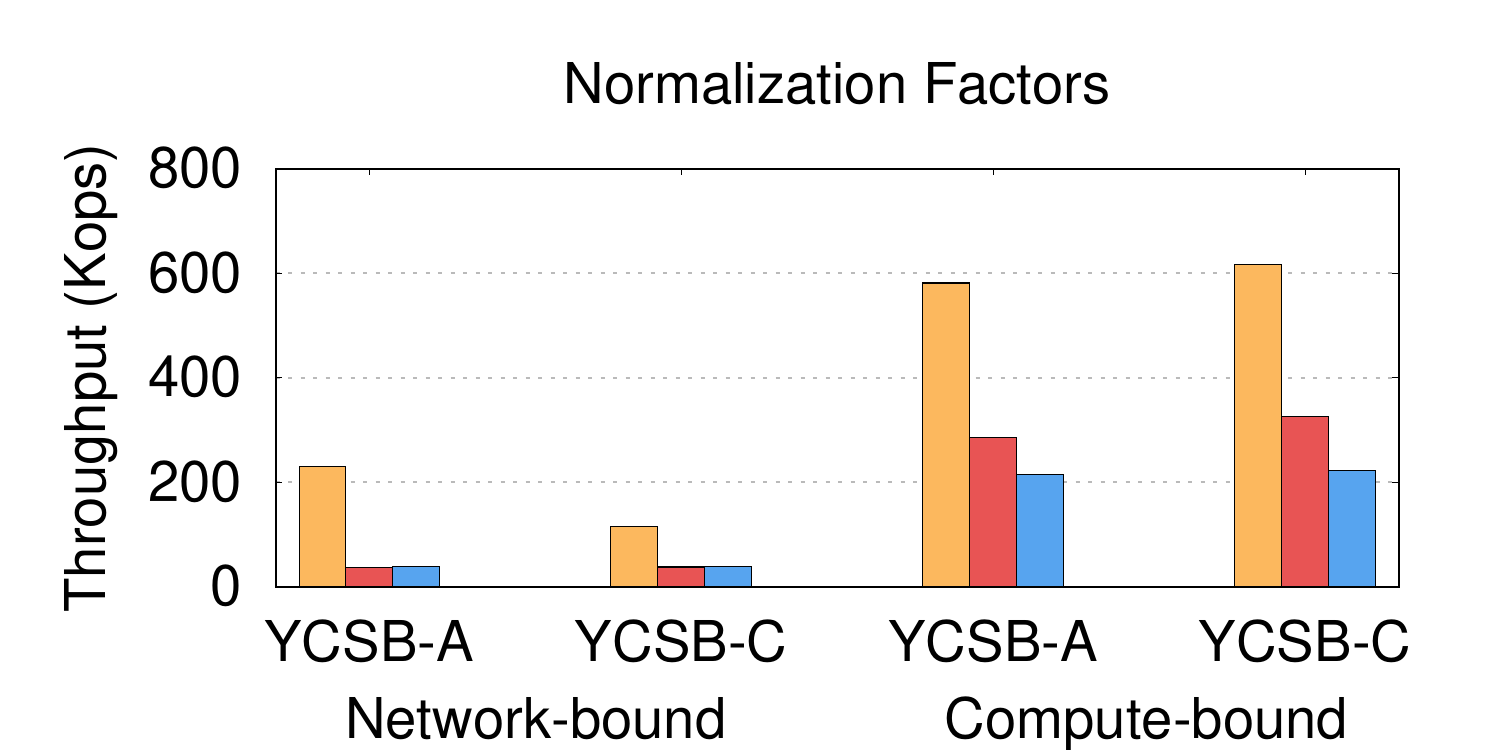}
  \caption{\textbf{Scalability properties of different systems when network bandwidth and compute are the bottleneck.} (left, middle) show system throughput normalized by throughput for a single physical proxy server, while (right) shows system throughput with a single physical proxy server. The Encryption-only lines for the network-bound and compute-bound cases overlap, since its throughput scales linearly in both cases. Since Pancake is centralized, it only has a single data point at $X=1$ for each of the cases, and these points overlap. See \S\ref{ssec:scalability-analysis} for details.
  }\label{fig:scalability}\vspace{-1.5em}
\end{figure*}

The following theorem establishes the security of \name under $\INDCDDA$:
\begin{theorem}[$\INDCDA$ Security]
  \label{theorem:indcda}
  Let $q \geq 0$ and $Q = q \cdot B$. Let $\dist_0, \estdist_0, \dist_1, \estdist_1$ be query distributions. For any $q$-query $\INDCDA$ adversary $\adv$ against \name there exist adversaries $\bdv$, $\cdv$, $\ddv_1$, $\ddv_2$ such that \\
  $$\AdvINDCDA_{\name}[(\adv)] \leq \AdvPRF_{F}[(\bdv)] + \AdvROR_{E}[(\cdv)]$$
  $$\myind\myind\myind\myind\myind\myind + \AdvDIST_{Q,\dist_0,\estdist_0}[(\ddv_1)] + \AdvDIST_{Q,\dist_1,\estdist_1}[(\ddv_2)]$$
  where $F$, $E$ are $PRF$, $AE$ schemes used by \name. Adversaries $\bdv, \cdv, \ddv_1,\ddv_2$ run in same time as $\adv$ with $Q$ queries.
\end{theorem}

\noindent
Our security proof stems from three key components:
\begin{itemize}[leftmargin=*, itemsep=0pt]
  \item Security of $E$ as a randomized authentication scheme applied over values and $F$ as a pseudorandom function applied over keys; this is rigorously analyzed in prior work~\cite{ggm,rogaway2006provable}.
  \item \maybeaddressed{PC Discussion}{Discuss their assumptions more clearly. Particularly why observing a subset of queries is representative of the whole (whereas prior work could see all queries). This might be somewhat reasonable at very high load, but what about for a system with variable load, where sometimes there are high loads and sometimes low loads. It is not clear why this assumption holds and authors are encouraged to dig more deeply into this.}
  \maybeaddressed{D}{the leader can estimate the new distribution from a small random sample of the batch of users' queries as effectively as prior work, even though prior work had the entire batch of users' queries. Why are these assumptions reasonable?}
  \maybeaddressed{B}{there's $\hat{\pi}$ which is an estimate of the true distribution $\pi$ and which your security proof assumes is 'good enough'; why can't I just build a simpler scale out scheme in which I lazily disseminate updates between proxies so that they can adjust their individual estimates $\hat{\pi}_{i}$  ... and then I can assume that those estimates are good enough. (I'm being a little facetious here; my main point is that there's some bounds on uncertainty that are parameters of a practical system, and you might want to explain that a bit more} 
  Our estimate $\estdist$ of the underlying distribution $\dist$ is sufficiently accurate. While this estimate may not be perfect, our security model only requires that $\estdist$ and $\dist$ be indistinguishable for a limited number of samples, which holds for estimators used in prior work~\cite{pancake} on real-world workloads~\cite{ycsb}. Since our design employs a single leader \li server to estimate the underlying distribution using the keys for all client queries (\S\ref{ssec:three-layer}) and employs the same estimators as prior work, its estimation is just as accurate.
  \item Accesses issued to the KV store reveal nothing about the underlying distribution $\dist$.
\end{itemize}

\noindent
To prove the third component, we introduce simulators to sequentialize the distributed execution of \name's query processing to make it compatible with our game-based definition (Figure~\ref{fig:indcda}). Specifically, we simulate $\Process$ function for \name by first generating the intermediate transcript, $\beta$, assuming no failures. We do so by (i) going layer by layer and executing processing logic at appropriate servers in \name, and (ii) incorporating the impact of network reordering across queries between layers. We then use a $\Transform$ simulator to capture the effect of failures and generate the final transcripts $\tau$ from $\beta$. We do so by recursively applying the effect of \liii server failure events in $\failures$ on the intermediate transcripts $\beta$ in the order that they occur.

Finally, we show that the final transcripts $\tau$ are independent of intermediate transcripts $\beta$, and then show that $\beta$ are independent of the underlying distribution $\dist$.
The first part holds since \name randomly shuffles buffered queries before replaying them post failure (\S\ref{sec:failures}) and failure recovery time in \name is short enough to not be visible to an external observer given our failure model (\S\ref{ssec:failures}) and as shown empirically in (\S\ref{ssec:failure-recovery}). The second part holds, since the underlying oblivious data access scheme~\cite{pancake} in \name generates uniform random queries (\S\ref{ssec:three-layer}) and network reorderings between layers are independent of $\dist$. 

Finally, to model dynamic distributions, we generalize the above definition to Indistinguishability under Chosen Dynamic Distribution and Failure Attack or $\INDCDDA$. This definition, along with the proof of \name's security under it, formal descriptions of our simulators, and the proof for independence of $\tau$ and $\dist$ are presented in\iffullversion  ~Appendix~\ref{app:proofs}\else~\cite{suppl}\fi.


\section{Evaluation}
\label{evaluation}
\label{sec:eval}

\name is implemented in $\sim 6$k lines of C++, using Thrift as the RPC library, AES-CBC-256 for encrypting values, HMAC-SHA-256 as our PRF, and Redis as the KV store. 

\paragraphb{Compared systems} 
We compare \name performance against two baselines. The first baseline is distributed, but encryption-only, that is, it encrypts data and client queries, but does not guarantee oblivious data access; here, client queries are randomly load balanced across stateless proxy servers that perform encryption/decryption and forward queries to the KV store. This baseline serves as an upper bound on the performance that can be achieved by any oblivious data access system (including \name). 
The second baseline is a centralized \textsc{Pancake}~\cite{pancake} proxy server. While this suffers from security and availability problems in the face of failures (\S\ref{ssec:beefy-single-proxy}), its performance serves as a reference point for understanding \name's scalability. 

\paragraphb{Experimental setup} 
We run our experiments on Amazon EC2. By default, we host the proxy instances across c5.4xlarge VMs with $16$ vCPUs ($8$ cores with $2$ threads per core), $32$ GB RAM and $10$Gbps network links.
In order to emulate a cloud KV store with practically infinite bandwidth, we use a single powerful VM, c5d.metal ($96$ vCPUs, $128$ GB RAM) with large network bandwidth ($25$ Gbps). Similar to prior work~\cite{pancake}, we emulate WAN access link bandwidth by throttling the bandwidth from each proxy server to the KV store server to $1$Gbps. The clients run on lightweight t3.2xlarge VMs ($8$ vCPUs, $32$GB RAM) in the same LAN. Both \pancake and \name use a batch size of $B=3$.

\paragraphb{Dataset and Workloads} 
\maybeaddressed{E}{The workloads used in the evaluation (1KiB values) seem too small for medical data. Even a single specialist report is probably larger than that in average. But the charts (presumably scans) consulted by the specialist would be orders of magnitude larger. How would larger payloads affect the performance?}
We use the standard YCSB benchmark~\cite{ycsb} to generate our dataset and workloads. The dataset comprises $1$ million KV pairs, with $8$B keys and $1$KB values. We use workloads A ($50$\% reads, $50$\% writes) and C ($100$\% reads) for our experiments. YCSB workloads perform accesses distributed according to the Zipfian distribution~\cite{ycsb}; unless otherwise stated, the skewness parameter for the Zipfian distribution in our experiments is set to the YCSB default of $0.99$ (that is, heavily skewed), which is representative of many real world workloads. We also perform sensitivity analysis against distribution skew.

\subsection{Scalability Analysis}
\label{ssec:scalability-analysis}

We now analyze \name's scalability with varying number of physical proxy servers under different workloads.

\paragraphb{Throughput scaling under bandwidth bottleneck} We study throughput scaling for \name by varying the number of physical proxy servers and comparing its performance against the baselines.  
For \name, $k$ physical proxy servers constitute $k$ chain-replicated \li instances with $\min(k, 3)$ replicas each, $k$ chain-replicated \lii instances with $\min(k, 3)$ replicas each, and $k$ unreplicated \liii instances (\ie, the system can tolerate up to $\min(k, 3) - 1$ failures). For the encryption-only baseline, a separate proxy instance is run on each physical proxy server, and the \pancake baseline always uses only one physical proxy server.

Figure~\ref{fig:scalability} shows the scalability results for two cases: one where the physical proxy servers are network-bound (solid lines), and another where they are compute-bound (broken lines). We begin with the former case; we see that \name throughput scales linearly with the number of physical proxy servers. Note that we normalize each system's throughput by its throughput with a single physical proxy server --- Figure~\ref{fig:scalability}~(right) shows normalization factors for each system, \ie, throughput with single physical proxy server. The red cross shows the throughput of the \pancake baseline ($38$ KOps): \name's distributed design enables linear throughput gains relative to \pancake via scaling. The insecure baseline also scales linearly due to random load-balancing across its proxy instances.
Since all proxy servers are network bound, \name incurs only a constant overhead (corresponding to the relative bandwidth increase due to the oblivious data access protocol) compared to the encryption-only baseline for all configurations as we scale the number of physical proxy servers. For the YCSB-C workload, the gap between \name and Encryption-only baseline throughput stems from the $3\times$ overhead imposed by the \pancake protocol for a batch size of $B=3$. 
For the YCSB-A workload, however, the encryption-only baseline throughput is $6\times$ higher than \name since it can exploit the bidirectional bandwidth to the KV store for $50\%$ reads and $50\%$ writes. \name, however, already issues a read followed by a write for every query, so it is unable to similarly exploit the bidirectional bandwidth. Since YCSB-A has equal proportion of read and write queries, this situation corresponds to the worst-case bandwidth increase ($6\times$) for \name relative to the encryption-only baseline.

\begin{figure*}
  \centering
  \includegraphics[width=0.32\linewidth]{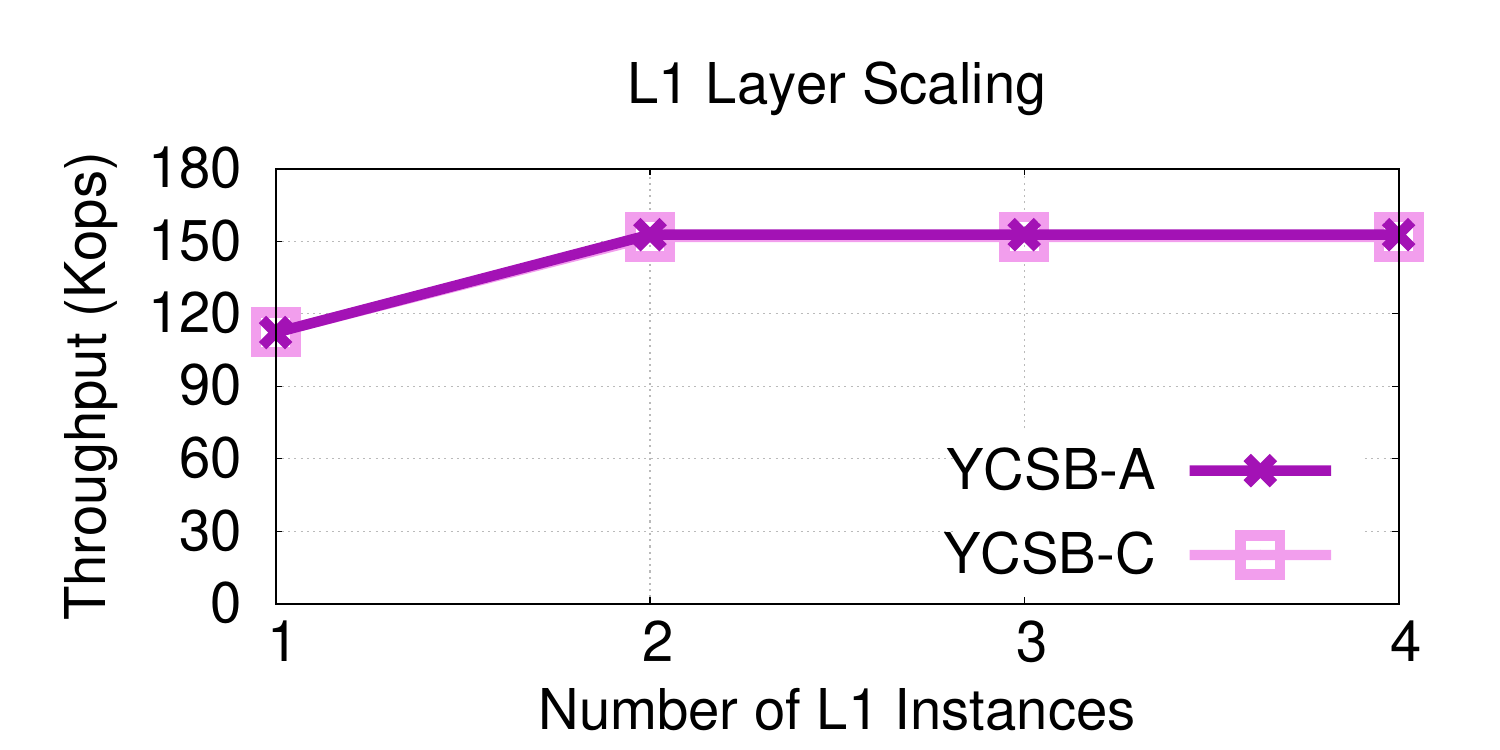}
  \includegraphics[width=0.32\linewidth]{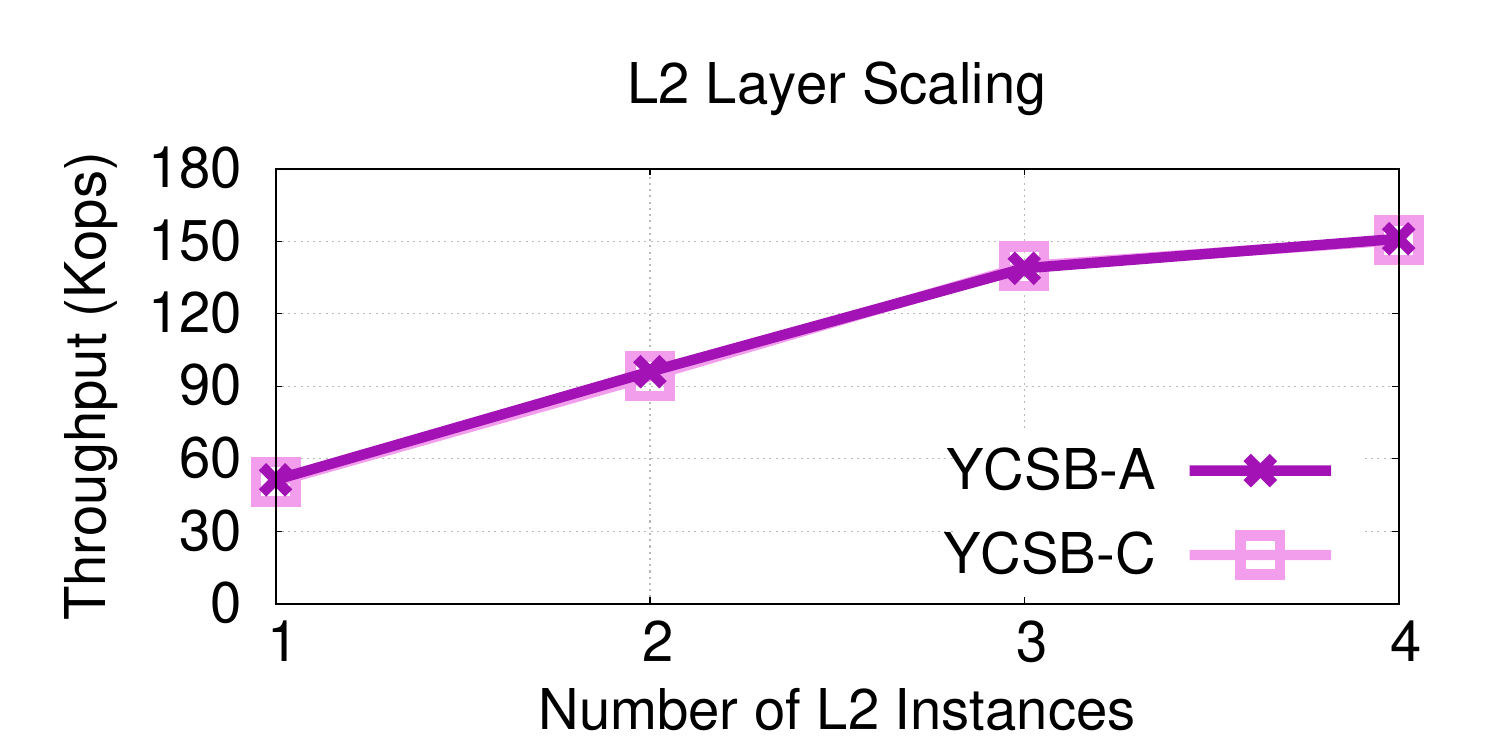}
  \includegraphics[width=0.32\linewidth]{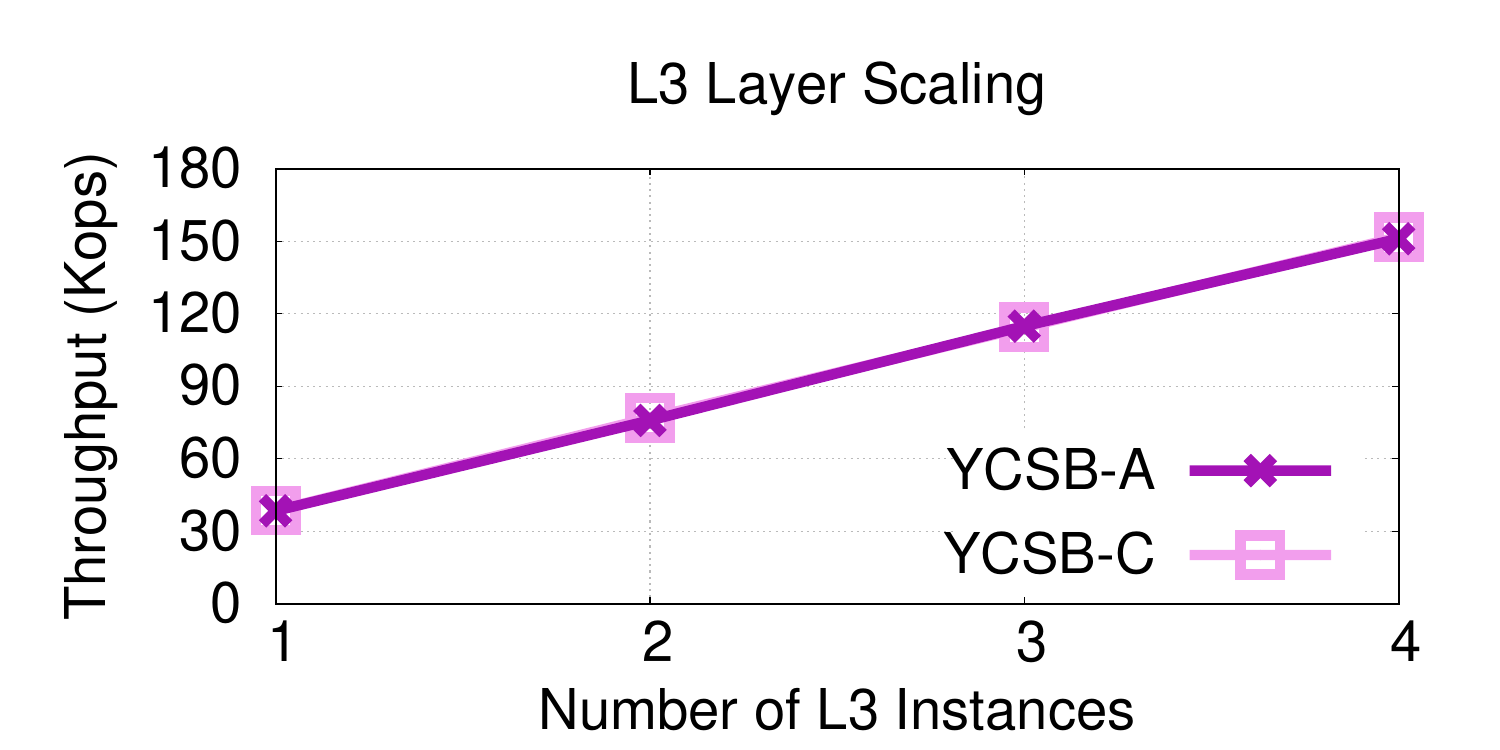}
  \vspace{-0.3em}
  \caption{\textbf{\name layer-wise scaling for YCSB workloads A and C.} See \S\ref{ssec:scalability-analysis} for details.}\label{fig:layerbottlenecks}\vspace{-1.5em}
\end{figure*}

\paragraphb{Throughput scaling under compute bottleneck}
We now analyze throughput scaling when the physical proxy servers are compute-bound: we re-run the same experiments as above, but using c5.metal EC2 VMs (96 vCPUs, 192GB RAM, 25Gbps network bandwidth) for all systems \emph{without} throttling the access link bandwidth to the KV store server. As the broken lines corresponding to the compute-bound case in Figure~\ref{fig:scalability} show, with a single physical proxy server \name achieves slightly lower throughput than \pancake for both workloads. This is because, under a compute bottleneck, \name incurs additional RPC processing overheads for communication between its layers. \name's throughput increases significantly with more physical proxy servers, achieving $3.4-3.6\times$ higher throughput with $4$ physical proxy servers. The increase in throughput is not perfectly linear, since workload skew results in load imbalance at the \lii layer. This effect is not observed for the network-bound case, since the network bandwidth between \liii instances and the KV store is bottlenecked before workload skew causes compute at the \lii layer to become bottlenecked. For the reminder of our evaluation, we use the network-bound setting as our default configuration.  

\begin{figure}
  \vspace{-1em}
  \centering
  \subfigure[{\name throughput is unaffected by access skew.}]{\label{fig:skew}\includegraphics[width=0.49\linewidth]{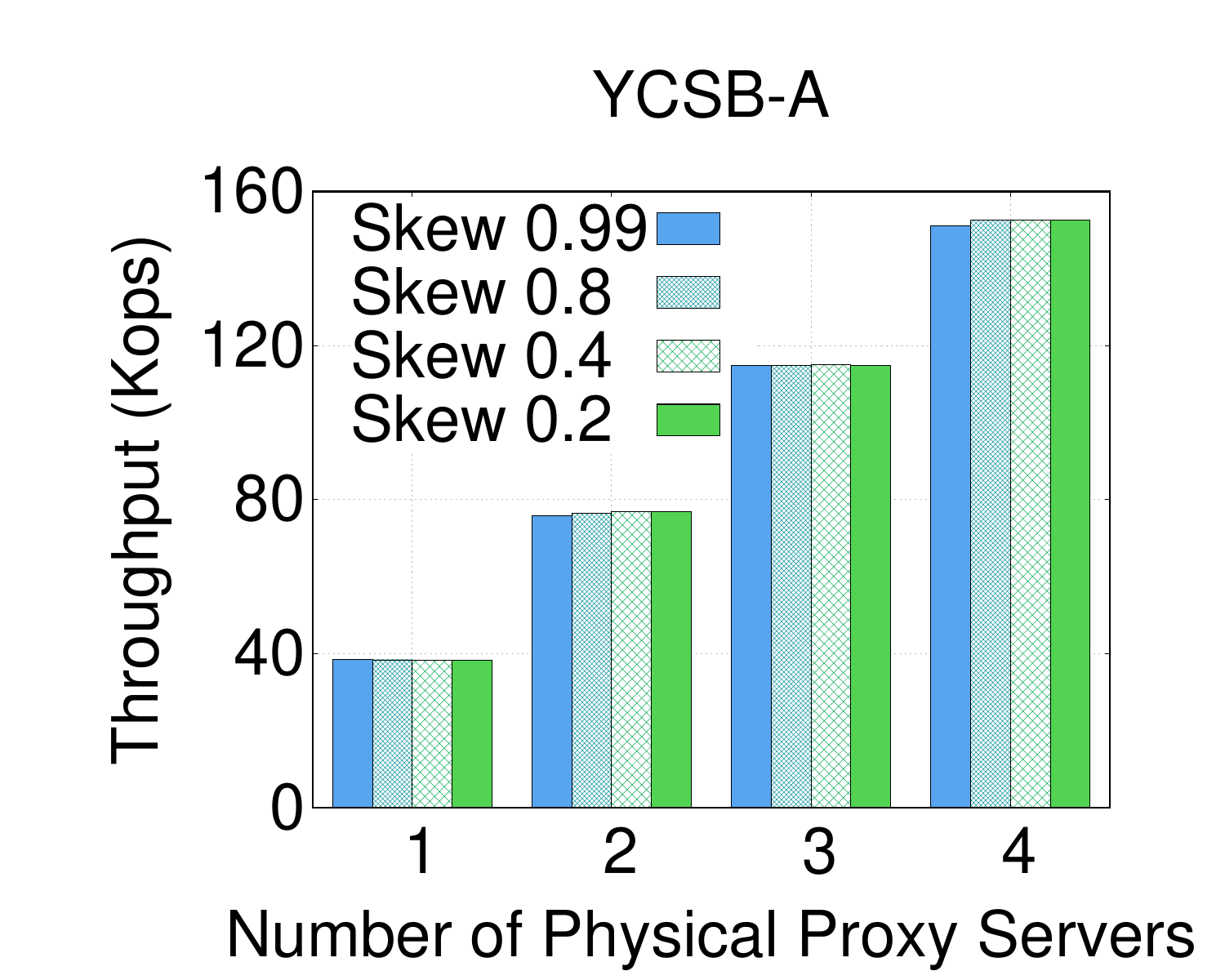}}
  \subfigure[{Query latency vs. number of physical proxy servers.}]{\label{fig:latency}\includegraphics[width=0.49\linewidth]{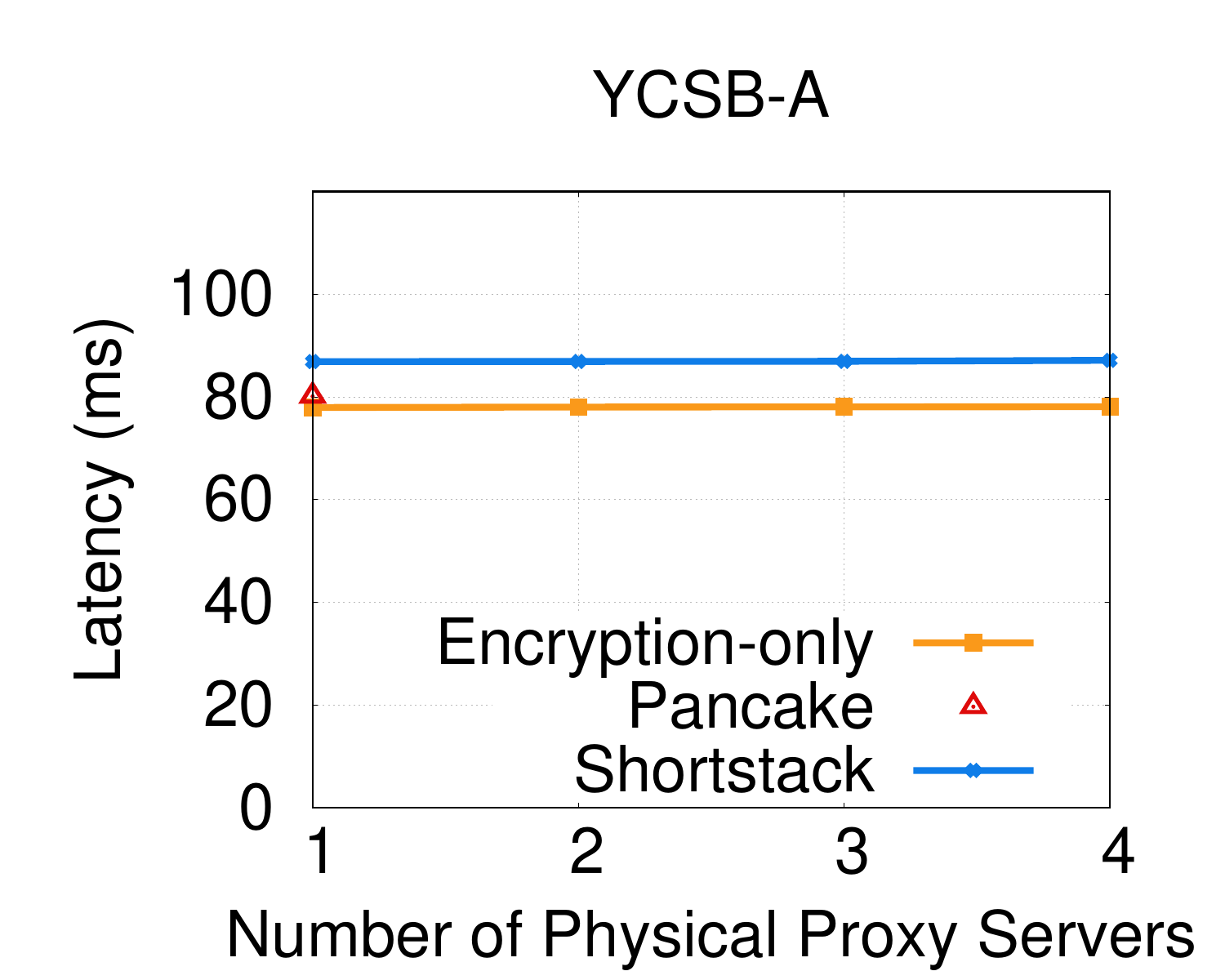}}\vspace{-.75em}
  \caption{\textbf{\name throughput scaling with varying skew (a) and \name latency overheads (b).} See \S\ref{ssec:scalability-analysis} for details.}\vspace{-2em}
\end{figure}

\begin{figure*}
  \centering
  \includegraphics[width=0.32\linewidth]{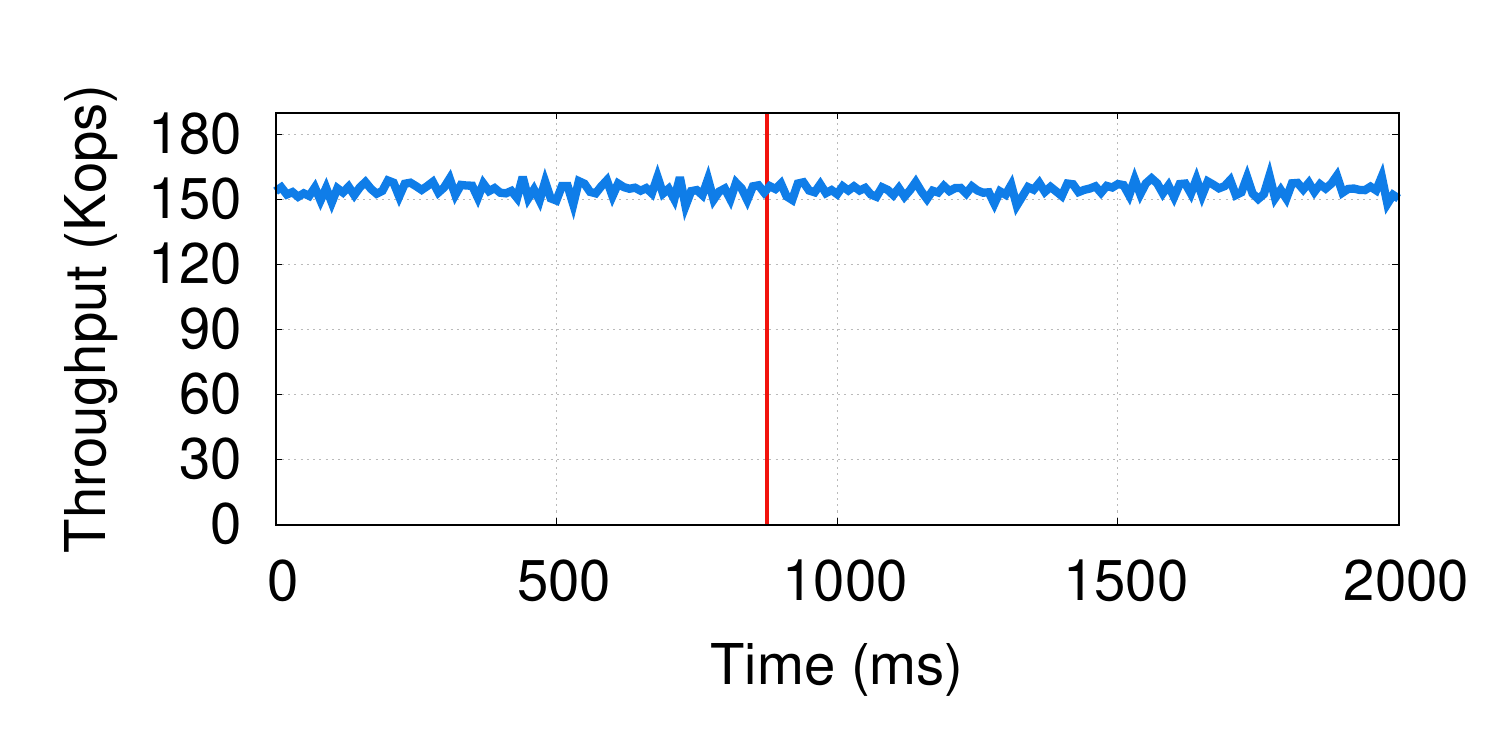}%
  \includegraphics[width=0.32\linewidth]{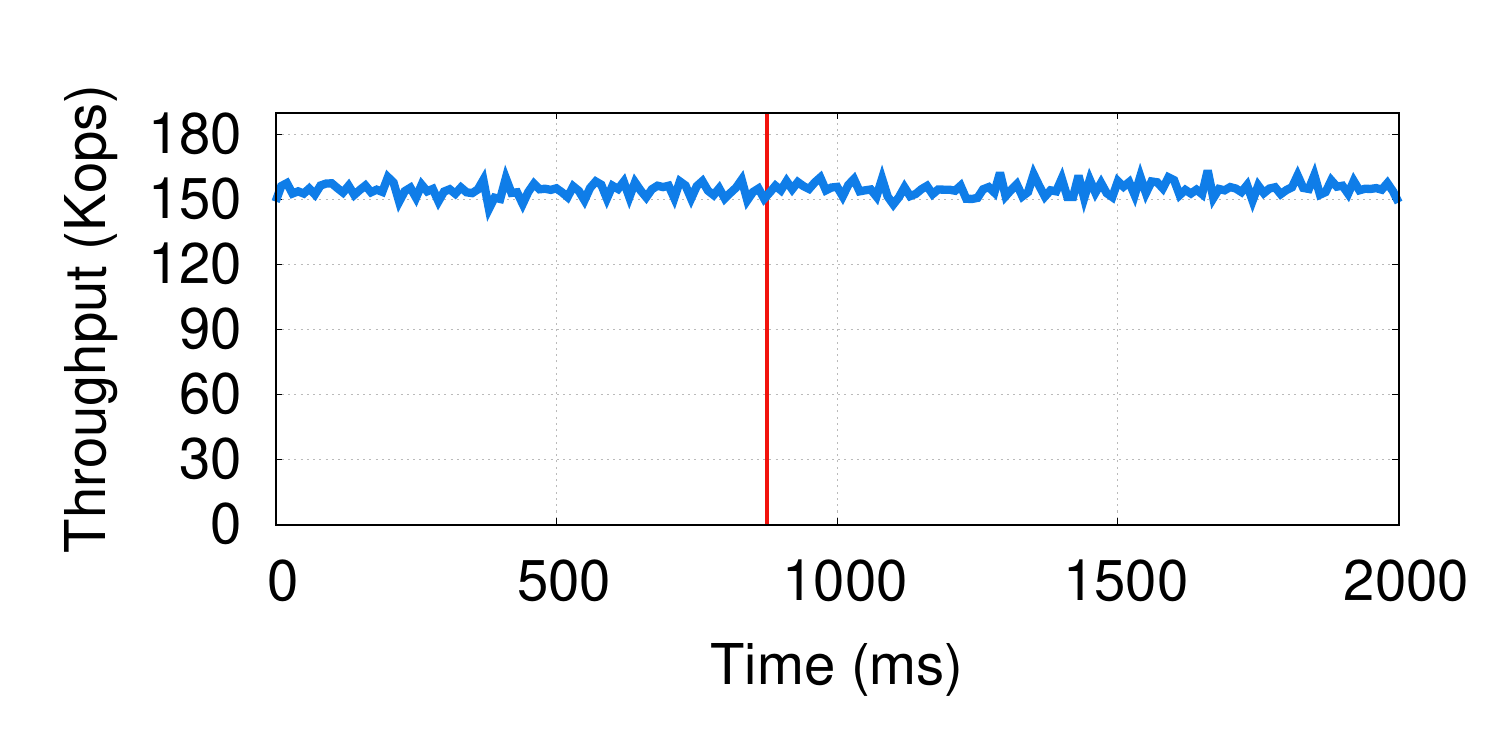}%
  \includegraphics[width=0.32\linewidth]{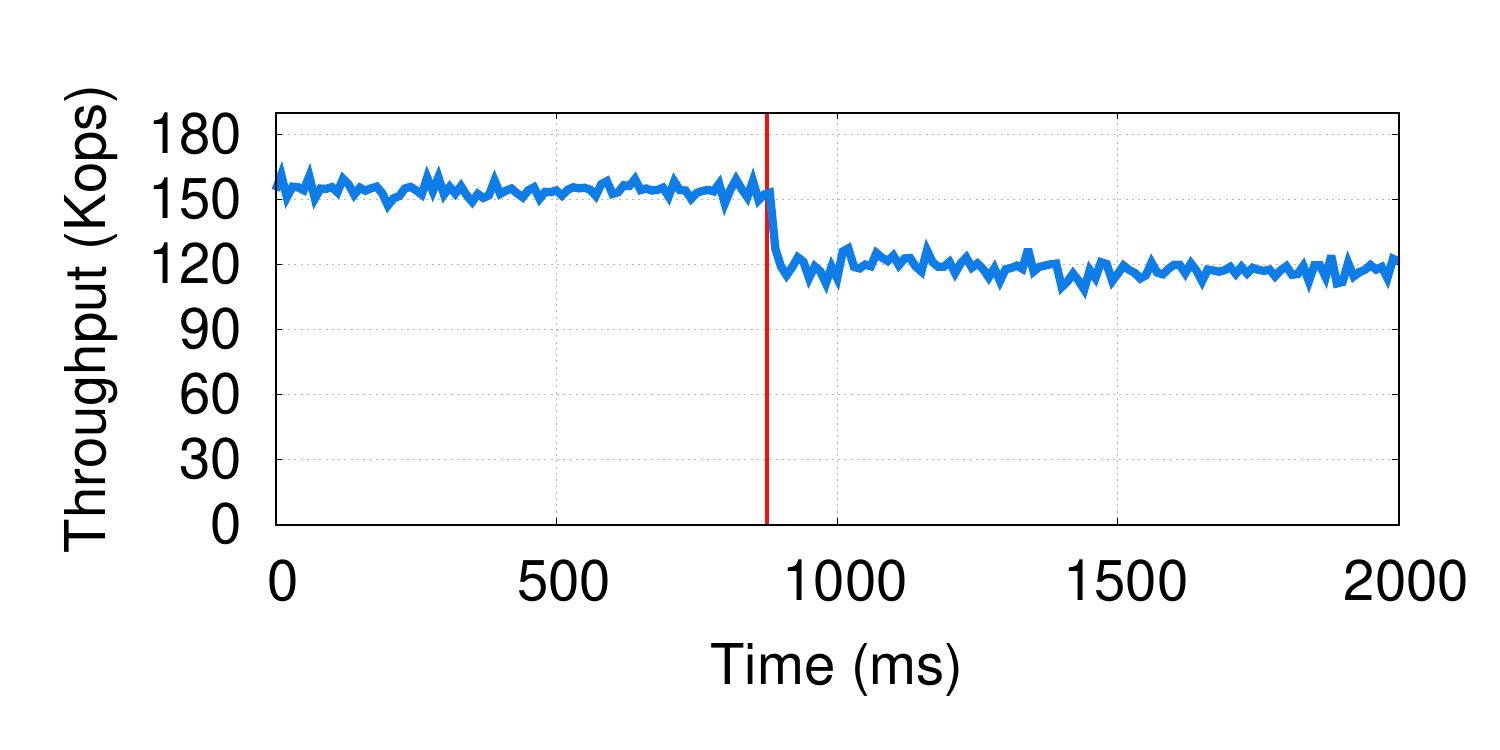}\vspace{-0.5em}
  \caption{\textbf{\name failure recovery for (left) \li, (middle) \lii, and (right) \liii failures.} See \S\ref{ssec:failure-recovery} for details.}\label{fig:failure}\vspace{-1.5em}
\end{figure*}

\paragraphb{Understanding per-layer scalability bottlenecks}
Our experiments in Figure~\ref{fig:scalability} scale up all layers of \name in equal proportions as the number of physical proxy servers are increased. To better understand \name's bottlenecks, we now study scalability on a per-layer basis.
Since each layer performs a different component of \pancake logic (\S\ref{sec:design}), varying the scale of each layer independently while keeping the scale of the other two layers fixed allows us to understand which step becomes a throughput bottleneck before the others. For this, we use a setup similar to Figure~\ref{fig:scalability}. To understand \li layer scalability, we fix the number of physical proxy servers to $4$, the number of replicated \lii instances and unreplicated \liii instances to the default ($4$), and vary the number of replicated \li instances from $1-4$. We perform similar experiments for the \lii and \liii layers as well.
Figure~\ref{fig:layerbottlenecks} shows the corresponding results for the YCSB-A and YCSB-C workloads. For the \li layer, throughput increases slightly from $X=1$ to $2$, beyond which it saturates, since \li is no longer the bottleneck. For the \lii layer, from $X=1$ to $3$ throughput increases, albeit non-linearly due to plaintext key-based partitioning --- while the number of plaintext keys handled by each \lii server is roughly equal, the number of replicas handled by them is skewed due to the skew in the YCSB workload. At $X=4$, the \lii layer is no longer the bottleneck. For the \liii layer, throughput scales linearly from $X=1$ to $X=4$ due to ciphertext key-based partitioning, with each \liii proxy handling roughly the same number of ciphertext keys.  

As expected, the bottlenecks are different at different \name layers. When all layers are sufficiently provisioned, \name is able to saturate the access link bandwidth between the \liii layer and the KV store. Reducing the number of \li and \lii proxy instances, however, leads to \textit{compute} becoming the bottleneck at the respective layers. One of the key contributors of compute overheads are serialization/deserialization for network queries. Finally, layer-wise scaling characteristics are similar for YCSB-C and YCSB-A workloads, as \updatecache processing in YCSB-A due to writes does not account for much of the compute overheads. 

\paragraphb{Throughput scaling with skew}
We evaluate \name scaling for workloads with different skew for a setup similar to Figure~\ref{fig:scalability}. We vary the skew parameter for YCSB's Zipf distribution from $0.2$ (close to uniform) to $0.99$ (heavy skew) to consider both extremes. We only show our results for YCSB-A in Figure~\ref{fig:skew}, since results for YCSB-C were similar. \name system throughput scales linearly regardless of skew, because the bottleneck in the end-to-end query execution is the access link bandwidth between the \liii layer and the KV store for all scales. Since the skew only affects processing at \lii layer (which is not the bottleneck), our throughput is independent of skew. While \name throughput scales linearly even for heavily skewed workloads, there could indeed be rare extreme-case scenarios where such would not be the case, \eg, if all popular plaintext keys get consistently hashed to a single \lii instance, resulting in a compute bottleneck at that instance.

\paragraphb{\name Latency overheads}
To quantify \name's latency overheads, we evaluate end-to-end query latency for varying number of physical proxy servers for compared systems using a setup similar to Figure~\ref{fig:scalability} with one change: we separate the KV store and physical proxy servers by the WAN. Figure~\ref{fig:latency} shows the results; again, we only show YCSB-A workload results, as YCSB-C results are similar. Independent of the scale, \name increases query latency by a modest $8$\% (additional $6.8$ms) compared to \pancake. This increase in latency is due to additional processing and network hops introduced by \name's multiple layers and chain replication within the \li and \lii layers. Nevertheless, these overheads are masked by the significantly larger WAN access latency.

\subsection{Failure Recovery}
\label{ssec:failure-recovery}

We now evaluate \name's ability to recover from failures and also validate our assumptions in proving \name security. We fix the number of physical proxy servers to $4$, the number of $3\times$-replicated \li, $3\times$-replicated \lii, and unreplicated \liii instances to $4$ each, and use the YCSB-A workload.
To understand the impact of failures on each layer independently, we fail one proxy instance in a particular layer by killing its associated process; for \li and \lii, we kill an arbitrary replica from one of the instances. We measure the instantaneous throughput of our system during each experiment at $10$ms granularity; when measured at finer-grained timescales, we found that the instantaneous throughput numbers were too noisy to discern any meaningful trends. 

Figure~\ref{fig:failure} shows the effect of failure at each layer on \name throughput. We find that failures in \li and \lii proxy chains do not cause any noticeable dip in the throughput, since \name can quickly recover from failures within $3$--$4$ ms --- much faster than the average query latency over WAN ($\sim 90$ms), and smaller than the typical variance in query latencies.
Hence, an adversary cannot reliably distinguish between a failure event and variations in instantaneous throughput due to noise caused by network delays, independent of the timescale at which measurements are done. This validates our assumption for \name security under failures discussed in \S\ref{sec:security} --- specifically, \li and \lii failures have an imperceptible impact on an adversary's observed access pattern to the KV store. Upon an \liii proxy failure, the throughput reduces by $25\%$ -- commensurate with the reduction in the bandwidth to the KV store server; however, since \liii layer partitions queries by ciphertext keys, it does not reveal any information about the client access patterns.


\section{Related Work}
\label{sec:related}
We now discuss the works most closely related to \name's goals of distributed, fault-tolerant, oblivious data access. 
ORAM~\cite{oram} approaches have been adapted to real world cloud storage~\cite{privatefs, oblivistore, shroud, taostore, curious, concuroram, dauterman2021snoopy}, with recent efforts enabling \textit{concurrency} and \textit{asynchrony}. Oblivious Parallel RAM (OPRAM)~\cite{boyle2016oblivious, chan2018perfectly, chan2017circuit, davies2020client, privatefs, shroud} permits multiple concurrent clients to query the storage, but requires cross-client coordination per-query (\eg, using oblivious aggregation~\cite{privatefs}) to ensure no two clients concurrently issue a query for the same data. This severely limits throughput scaling under high query traffic due to compute bottlenecks.

CURIOUS~\cite{curious} and TaoStore~\cite{taostore} employ a centralized proxy model, 
but permit client parallelism via \emph{asynchrony}. Since each operation 
requires updates to per-plaintext key proxy state for multiple random KV pairs, extending their 
design to a distributed and secure one is challenging. The latest in this line of work,
ConcurORAM~\cite{concuroram} and Snoopy~\cite{dauterman2021snoopy}, permit multiple parallel clients to query a cloud-hosted ORAM
\textit{without} inter-client or proxy based coordination. ConcurORAM achieves this by offloading much of the synchronization to the cloud, which not only requires non-trivial changes to cloud storage, but also limits system throughput under high load. Concurrent to our work, Snoopy builds a distributed oblivious data access system (for ORAM-based designs); however, Snoopy does not prove security for scenarios where servers can fail. In any case, \name and Snoopy offer the same trade-offs as discussed in \cite{pancake}---Snoopy can handle active adversaries, but also incurs significantly higher overheads relative to \name. 
Prior work~\cite{pancake} has empirically shown that state-of-the-art single proxy ORAM schemes achieve $220\times$ lower throughput than \pancake for the same workloads as in our evaluation. Since \name can scale \pancake's throughput linearly (\S\ref{sec:eval}) with number of proxy servers, even if one could design a distributed ORAM system that scales near-perfectly with number of proxy servers, the maximum achievable throughput would still be orders of magnitude lower than \name.

\section{Conclusion}
Existing systems for oblivious data access rely on a centralized, stateful, proxy to coordinate queries between applications and the storage server. We have demonstrated that, in failure-prone deployment, such systems can suffer from security violations, long periods of unavailability and/or scalability limits. Our core contribution is \name, a distributed, fault-tolerant and scalable system for oblivious data access. Using a novel layered architecture, \name achieves the classical obliviousness guarantee---access patterns observed by the adversary being independent of the input---even under a powerful passive persistent adversary that can force failure of arbitrary (bounded-sized) subset of proxy servers at arbitrary times. We also introduce a security model to study oblivious data access with distributed, failure-prone, servers.

\section*{Acknowledgements}
We would like to thank our shepherd, Alex C. Snoeren, and the anonymous OSDI reviewers for their insightful feedback. We would also like to thank Thomas Ristenpart for many useful discussions during this work. This research was supported in part by NSF awards 2054957, 2047220, 2118851, 1704742, Faculty Research Awards from Google and NetApp, and an IC3 fellowship thanks to IC3 industry partners.
\iffullversion
  
  \bibliographystyle{unsrt}
  \bibliography{bib/citations}
  \appendix
  \iffullversion
\else
\customlabel{ssec:three-layer}{4.2}
\customlabel{sec:failures}{4.3}
\customlabel{ssec:failures}{4.3}
\customlabel{sec:eval}{6}
\customlabel{ssec:failure-recovery}{6.2}
\customlabel{sec:security}{5}
\customlabel{fig:l1_logic}{8}
\customlabel{ssec:dynamic_dist}{4.4}
\fi

\section{Security Proofs}
\label{app:proofs}

\paragraphb{Model Assumptions} In all following cases, we assume that the adversary has access to \name's encryption of the KV store $\DB'$, and a transcript $\tau$ generated by \name in response to $\numqueries$ queries drawn from a possibly time varying distribution. We assume network latencies between \li, \lii and \liii proxy servers to be stable, with variance indistinguishable from noise. We do not hide the timing of client accesses. 

For the case of static distributions, we let the adversary select up to $f$ proxy servers that are failed. We also note that the failure repair (\ie, detecting and eliminating the failed node, and ensuring the remaining nodes can resume operation) in chain replication occurs in bounded time $T_{\text{repair}}$ -- small enough that there is no \emph{detectable} dip in the throughput. This is consistent with our failure model described in \S\ref{sec:failures} and confirmed in our experiments in \S\ref{ssec:failure-recovery}.

For the case of dynamic distributions, in addition to the assumptions stated for static distributions, we assume \name \li leader can detect distribution changes instantaneously (as in~\cite{pancake}). Since \name's distribution change detection is the same as in \pancake~\cite{pancake}, and occurs at a single leader \li server that observes keys of all client queries (\S\ref{ssec:three-layer}, \S\ref{ssec:dynamic_dist}), the detection is just as accurate as in \pancake.
Further, we assume the distribution change is propagated to all proxy servers instantaneously. If detection and/or propagation takes finite time, \name --- as well as \pancake --- experiences a short period of bias in distribution proportional to the difference between $\estdist$ and $\newestdist$. However, prior work~\cite{pancake} shows that exploiting this bias to learn anything meaningful about the underlying distribution is challenging.

We now provide some technical preliminaries, then prove \name security theorems.

\paragraphb{Preliminaries} Let $\AdvPRF_{F}(\bdv)$ be the advantage $\bdv$ has in distinguishing
the pseudo-random function (PRF) $F$ from a random oracle. Let $E$ be an \textit{authenticated encryption
  with associated data} (AEAD) scheme characterized by the triple of
algorithms: ($\KeyGen$, $\Enc$, $\Dec$). We define $\AdvROR_{E}(\cdv)$ to be the
adversary's advantage in distinguishing $E.\Enc$ from a random oracle.
Next, we define $\AdvDIST_{q,\dist,\estdist}(\ddv)$ as the advantage an
adversary $\ddv$ has in computationally distinguishing between the actual and estimated distributions
$\dist$ and $\estdist$ using $q$ samples.

\paragraphb{Security analysis for static distributions} We define the chance of success of an adversary $\adv$ in attacking an encrypted KV store $\EDB$ via frequency analysis, as the advantage it has in guessing the right underlying distribution in the game
$\INDCDA$ (Figure~\ref{fig:indcda-appendix}) as follows:
\begin{align*}
  \AdvINDCDA_{\EDB}(\advA) =& \; \lvert\Pr[\INDCDA^{\adv}_{1,q,S,f} \Rightarrow 1]\\
                           & \hspace*{3em}-\ \Pr[\INDCDA^{\adv}_{0,q,S,f} \Rightarrow 1]\rvert\,.
\end{align*}%
\begin{figure}[t]
  \centering
  \subfigure[\INDCDA]{
    \fpage{.21}{\scriptsize
    \procedurev{$\INDCDA^{\adv}_{b,q,S,f,\dist_0,\hat\dist_0,\dist_1,\hat\dist_1}$} \\
    $\DB, \failures, \stadv  \sample \adv_1(f, S)$ \\
    $(\DB',\chains,\delta) \sample \Init(\hat{\dist_b}, \DB, S, f)$ \\
    For $i$ in $1$ to $q$: \\
    \myind $w \sample \dist_b$ \\
    \myind $W \gets W \cup \{w\}$ \\
    $\tau_{1}, \tau_{2}, ... \gets \Process(W,\chains,\failures,\DB',\delta)$ \\
    $b' \sample \adv_3(\stadv, \DB', \tau_{1}, \tau_{2}, ...)$ \\
    \textbf{return} $b'$
    }
    \label{fig:indcda-appendix}
  }%
  \subfigure[\INDCDDA]{
    \fpage{.23}{\scriptsize
    \procedurev{$\INDCDDA^{\adv}_{b,q,S,f,\dist_0, \estdist_0, \dist'_0, \estdist'_0, \dist_1, \estdist_1, \dist'_1, \estdist'_1}$} \\
    $\DB, c, \failures, \stadv  \sample \adv_1(f, S)$ \\
    $(\DB',\chains,\delta) \sample \Init(\estdist_b, \DB, S, f)$ \\
    For $i$ in $1$ to $c$: \\
   \myind $w \sample \dist_b$ \\
   \myind $W \gets W \cup \{w\}$ \\
    $\tau_{1}, \tau_{2}, ... \gets \Process(W,\chains,X,\DB', \delta)$\\
    For $i$ in $c$ to $q$: \\
    \myind $w \sample \dist'_b$ \\
    \myind $W' \gets W' \cup \{w\}$ \\
    $\tau'_{1}, \tau'_{2}, ... \gets \Process(W',\chains,X,\DB', \delta, \newestdist_b) $ \\
    $b' \sample \adv_3(\stadv, \DB', \tau_{1} ,\tau'_{1}, \tau_{2}, \tau'_{2}, ...)$ \\
    return $b'$
    }
    \label{fig:indcdda-appendix}
  }
  \caption{\textbf{Security games in \name.}}\vspace{-1em}
\end{figure}
We already discussed the game description in \S\ref{sec:security} of the main paper --- we now focus on detailing the simulators employed in \name to translate the distributed, fault-tolerant query execution in \name to a sequentialized version compatible with the $\INDCDA$ game.

\begin{figure}[h]
  \centering
  \fpage{.38}{
  \procedurev{$\Process(W,\chains,\failures,\DB',\delta)$} \\
  $C_{\li},C_{\lii},C_{\liii} \gets \chains$ \\
  $k_F \sample \mathcal{K};k_{AE} \sample \mathcal{K};$ \\
  For $(k,v)$ in $W$: \\
  \myind $s_{\li} \sample C_{\li}$ \\
  \myind $s_{\li}$.ProcessQuery$(k,v)$ \\
  For $s_{\lii}$ in $C_{\lii}$: \\
  \myind $\Interleave(s_{\lii}.\Queue)$ \\
  \myind While $s_{\lii}.\Queue$ is non-empty: \\
  \myind \myind $s_{\lii}$.Process() \\
  For $s_{\liii}$ in $C_{\liii}$: \\
  \myind $\Interleave(s_{\liii}.\Queue)$ \\
  \myind While $s_{\liii}.\Queue$ is non-empty: \\
  \myind \myind $(\kw', \val) \sample s_{\liii}$.Process($\delta$)\\
  \myind \myind $\beta_{s_{\liii}} \gets \beta_{s_{\liii}} \cup {(\kw', \val)}$ \\
  $\tau_{1}, \tau_{2}, ..., \tau_{|C_{\liii}|} \sample \Transform (\failures, C_{\liii}, \beta_{1}, \beta_{2}, ..., \beta_{|C_{\liii}|})$ \\
  \textbf{return} $\tau_{1}, \tau_{2}, ..., \tau_{|C_{\liii}|}$ \\
  }
\caption{\name $\Process$ algorithm.}\label{fig:Process_algo}\vspace{-1em}
\end{figure}

\paragraphc{Simulating \name query processing logic:} Figure~\ref{fig:Process_algo} shows the $\Process$ function for \name; note that the function \textit{simulates} the behavior of \name on the query vector $W$. 

It works by first generating the intermediate transcripts $\beta$ assuming no failures, going layer by layer and executing processing logic at appropriate servers in \name\xspace layers (\li, \lii, \liii). Specifically, in the first loop, queries are first processed at a randomly chosen chain $s_{\li}$ in $C_{\li}$. In the second loop, for each chain $s_{\lii}$ in $\lii$ layer, the enqueued queries are first shuffled using the $\Interleave$ function to simulate the effect of network-induced inter-leavings between servers in $\li$ layer and $s_{\lii}$, and then processed according to \lii processing logic. Note that the shuffling done by the $\Interleave$ function does not depend on the contents of the queries. Similarly, in the third loop, for each server $s_{\liii}$ in $\liii$ layer, queries are first shuffled using the $\Interleave$ function to simulate network reorderings, and then processed according to \liii $\Process$ logic. The generated queries are subsequently added to the transcript $\beta_{s_{\liii}}$. Finally, $\Process$ then uses a $\Transform$ simulator to simulate the effect of failures and generate the final transcripts $\tau$ from $\beta$, as we describe next.

\begin{figure}[h]
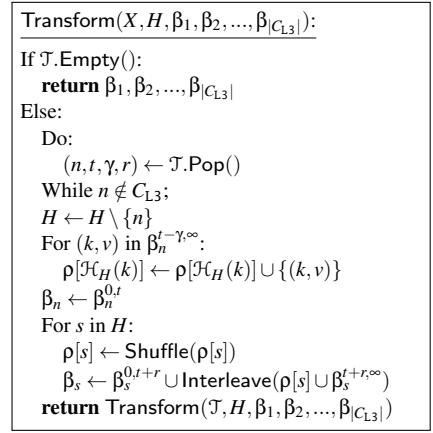

  \centering
  \fpage{.3}{
    \procedurev{$\Transform (X, H, \beta_{1}, \beta_{2}, ..., \beta_{|C_{\liii}|})$} \\
    If $\failures.\Empty()$:\\
    \myind \textbf{return} $\beta_{1}, \beta_{2}, ..., \beta_{|C_{\liii}|}$ \\
    Else: \\
    \myind Do:\\
    \myind \myind $(n,t,\gamma,r) \gets \failures.\Pop()$\\
    \myind While $n \notin C_{\liii}$;\\
    \myind $H \gets H \setminus \{n\} $ \\
    \myind For $(k, v)$ in $\beta_n^{t-\gamma, \infty}$:\\
    \myind \myind $\Retries[\mathcal{H}_{H}(k)] \gets \Retries[\mathcal{H}_{H}(k)] \cup \{(k, v)\}$\\
    \myind $\beta_{n} \gets \beta_{n}^{0, t}$\\
    \myind For $s$ in $H$:\\
    \myind \myind $\Retries[s] \gets \Shuffle(\Retries[s])$\\
    \myind \myind $\beta_{s} \gets \beta_{s}^{0, t+r} \cup \Interleave(\Retries[s] \cup \beta_s^{t+r, \infty})$\\
    \myind \textbf{return} $\Transform(\failures, H, \beta_{1}, \beta_{2}, ..., \beta_{|C_{\liii}|})$
  }
  \caption{\name~$\Transform$~simulator for failures.}\label{fig:simulator}\vspace{-1em}
\end{figure}

\paragraphc{Failure simulation using $\Transform$:} As noted in \S\ref{ssec:failures} of the main paper, \name handles the failures in layers \li and \lii transparently using chain replication. Also, recall that failure repairs in \name (\ie, detecting and eliminating the failed node, and ensuring the remaining nodes can resume operation) occurs in bounded time which is small enough to render failures at the \li and \lii layers undetectable to an adversary (consistent with our failure model in \S\ref{sec:failures} and empirival validation in \S\ref{ssec:failure-recovery}), leaving only the handling of \liii failures. We model this using the $\Transform$ simulator --- a recursive algorithm that repeatedly applies the effect of \liii server failure events in $\failures$ on the intermediate transcripts $\beta$, in the order that the failures occur. We detail the operation of $\Transform$ next; note that we use the term $\beta_n^{t_1,t_2}$ to denote the subset of queries from the intermediate transcript $\beta$ enqueued at server $n$ in the time interval $[t_1, t_2]$.

In the base case with no \liii server failures (\ie, $\failures$ is empty), the output transcript $\tau$ is the same as the input transcript $\beta$. Otherwise, the algorithm dequeues failure events from the queue $\failures$, ignoring non-\liii failures. When an \liii failure event $e = (n, t, \gamma, r)$ is encountered, $\Transform$ takes the following actions to simulate its behavior. First, $n$ is removed from the set of healthy servers $H$ at the \liii layer. Since $t-\gamma$ is the time at which the last query was acknowledged at server $n$ prior to failure, $\beta_n^{t-\gamma,\infty}$ denotes all the queries that were either unacknowledged and in-flight to the server, or unsent queries after the failure occurred at $n$. The second step taken by $\Transform$ for simulating a failure event is therefore to repartitions all queries in $\beta_n^{t-\gamma,\infty}$ across the remaining healthy servers using the consistent hash function $\mathcal{H}_H$, storing the repartitioned queries in $\Retries$. Third, the server $n$'s output transcript is truncated to $\beta_n^{0, t}$, the set of all the queries from the beginning of $\Process$ simulation to the time $t$ of the failure. Next, the repartitioned queries $\Retries$ are appended to the transcripts of the corresponding healthy servers, after randomly shuffling them, and interleaving with the remaining queries $\beta_s^{t+r, \infty}$ (where $r$ is the failure recovery time) at the healthy server to simulate network reorderings. The random shuffle is necessary for the secure replay of queries on \liii failure, as described in \S\ref{ssec:failures} of the main paper. It makes the order of the replayed queries independent of the original queries, hiding any correlation which the adversary could otherwise exploit to learn about the sharding scheme at \lii layer and thus learning distribution sensitive information. Finally, the $\Transform$ function enters recursion with the updated transcripts and set of healthy servers to process the remaining set of failure events.

\paragraphc{Security guarantees due to $\ninj$:} The underlying oblivious data access scheme, \pancake~\cite{pancake} ($\ninj$) used in \name provides the following security guarantee, which we leverage to prove \name security: an adversary cannot distinguish the output transcript generated by repeated application of $\ninj.\Batch()$ on a sequence of $q$ queries, from a sequence of $q\cdot B$ uniformly distributed accesses to random bit strings, given that the PRF and encryption functions are replaced by truly random functions and $\ninj$'s estimate of the underlying distributions ($\estdist$) is the same as the distribution ($\dist$) from which queries are sampled. Less formally, the output transcript generated by $\ninj$ is independent of the underlying access distribution $\dist$. This guarantee is provided by \pancake.

We are now ready to prove \name's security under \INDCDA.
\begin{theorem}[$\INDCDA$ Security]
  Let $q \geq 0$ and $Q = q \cdot B$. Let $\dist_0, \estdist_0, \dist_1, \estdist_1$ be query distributions. For any $q$-query $\INDCDA$ adversary $\adv$ against \name there exist adversaries $\bdv$, $\cdv$, $\ddv_1$, $\ddv_2$ such that \\
  $$\AdvINDCDA_{\name}(\adv) \leq \AdvPRF_{F}(\bdv) + \AdvROR_{E}(\cdv)$$
  $$\myind\myind\myind\myind\myind\myind + \AdvDIST_{Q,\dist_0,\estdist_0}(\ddv_1) + \AdvDIST_{Q,\dist_1,\estdist_1}(\ddv_2)$$
  where $F$ and $E$ are the $PRF$ and $AE$ schemes used by \name. Adversaries $\bdv, \cdv, \ddv_1,\ddv_2$, each run in same time as $\adv$, and make $Q$ queries.
\end{theorem}

\begin{proof}
We prove the above theorem using a sequence of standard cryptographic game transitions and reductions. We first substitute $\Init$ and $\Process$ in game $\INDCDA$ with the algorithms in \name (see Figure~\ref{fig:l1_logic} of the main paper and Figure~\ref{fig:Process_algo}). Next, we replace the PRF $F$ and the encryption scheme $E$ with a truly random function and replace $\estdist_0$ and $\estdist_1$ with $\dist_0$ and $\dist_1$ respectively to obtain game $G$. Clearly, the difference between the success of an adversary $\advA$ in the game $G$ and $\INDCDA$ can be upper bounded by the sum of advantages of (i) the PRF adversary $\advB$ against the PRF scheme $F$, (ii) the real-or-random adversary $\advC$ against the encryption scheme $E$, (iii) the distribution adversary $\advD_1$ in computationally distinguishing between $\estdist_0$ and $\dist_0$, and, (iv) the distribution adversary $\advD_2$ in computationally distinguishing between $\estdist_1$ and $\dist_1$:
$$\left|\underset{b \sim \{0,1\}}{\Pr}[\INDCDA^{\adv}_{b,q,S,f} \Rightarrow b] - {\Pr}[G^{\adv}_{b,q,S,f} \Rightarrow b]\right| \leq$$
$$\AdvPRF_{F}(\bdv) + \AdvROR_{E}(\cdv) + \AdvDIST_{Q,\dist_0,\estdist_0}(\ddv_1) + \AdvDIST_{Q,\dist_1,\estdist_1}(\ddv_2)$$
We now argue that the underlying distribution (determined by bit $b$) is indistinguishable
to the adversary under game $G$. Formally,
\begin{align}
  {\Pr}[G^{\adv}_{0,q,S,f} \Rightarrow 1] = {\Pr}[G^{\adv}_{1,q,S,f} \Rightarrow 1]\label{eq:ind_g}
\end{align}
\noindent
We first make the important observation that the output of the $\Transform$ function $\tau_1, \tau_2 ...$ is independent of the underlying distribution if the input transcripts $\beta_1, \beta_2 ...$ are independent of the underlying distribution. To see why this property holds, note that the only other inputs to the function are the set of failure events $\failures$, and the set of healthy servers $H$ both of which are assumed to be independent of the underlying distribution. Therefore, to show Eq.~\ref{eq:ind_g} holds, it suffices to show that $\beta_{1}, \beta_{2}, ..., \beta_{|C_{\liii}|}$ are independent of the underlying distribution.

  The labels and values in $G$ are random strings for both $b=0$ and $b=1$. Since, each access is independent, it is sufficient to prove that the label in a given access is independent of the underlying distribution $\dist$. For a particular query in $\beta_i$, let $\zeta_i$ be the random variable denoting the replica being accessed. We prove that for any replica $(k,j)$ handled by $s_i$ in $\liii$ layer:
$$\Pr[\zeta_i = (k,j)] = \frac{1}{n'_i}\; .$$ 
\noindent
where $n'_i$ is the total number of labels handled by server $s_i$ in $\liii$ layer in case of no failures. Note that $n'_i$ is independent of the underlying distribution, since the number of labels handled by an $\liii$ server depends only on hash partitioning of ciphertext labels across $\liii$ servers.

To see why the above equation holds, let $L$ be the random variable representing the proxy in \li layer that receives the query from the client and forwards it
to the appropriate server in \lii layer. Then,
\begin{align*}
  \Pr[\zeta_i = (k,j)] &= \sum_{l \in S_\li} \Pr[\zeta_i = (k,j) \cap L = l] \\
                     &= \sum_{l \in S_\li} \Pr[\zeta_i = (k,j) | L = l] \cdot \Pr[L = l]
\end{align*}
\noindent
Since client queries are \textit{randomly} load balanced across $\li$ proxy servers, $P[L = l] = 1 / |S_\li|$. Therefore,
\begin{align*}
  P[\zeta_i = (k,j)] &= \sum\limits_{l \in S_\li}\frac{\Pr[\zeta_i = (k,j) | L = l]}{|S_\li|}
\end{align*}
\noindent
Moreover, since each $\li$ proxy server independently applies oblivious data access scheme $\ninj$, we have:
$$\Pr[\zeta_i = (k,j) | L = l] = \frac{1}{n'_i}$$
\noindent
and therefore,
$$P[\zeta_i = (k,j)] = \sum\limits_{l \in S_\li}\frac{1}{n'_i*|S_{\li}|} = \frac{1}{n'_i}$$
\noindent
completing the proof.  
\end{proof}

\paragraphb{Security analysis for dynamic distributions} We introduce a new game, Indistinguishability under Chosen Dynamic Distribution and Failure Attack or $\INDCDDA$ (Figure~\ref{fig:indcdda-appendix}) for reasoning about the security of \name under dynamic distributions; it is a generalization of $\INDCDA$ with two major changes:
\begin{itemize}[itemsep=0pt, leftmargin=*]
  \item The adversary additionally outputs a query index $c$ at which the distribution changes. 
  \item The distributed proxy scheme's $\Process$ function takes an additional argument for the new distribution, $\newdist_b$. In \name, we use $\newdist_b$ to recalculate the weights $\delta$, and perform replica swapping at $\li$ (\S\ref{ssec:dynamic_dist} in the main paper).
\end{itemize}

\paragraphc{Security guarantees due to $\ninj$:} We rely on the guarantees provided by the oblivious data access scheme, \pancake~\cite{pancake} ($\ninj$) to prove the security guarantees for \name. Specifically, the oblivious data access scheme $\ninj$ guarantees that no distribution sensitive information is revealed during and after the transition to the new distribution $\newestdist$, using the replica swapping approach. As with the static distribution setting, this guarantee is provided by the oblivious data access scheme used in our implementation~\cite{pancake}.

We define the probability of success for an adversary $\adv$ in attacking $\EDB$ as its advantage in guessing the right underlying distributions in the game $\INDCDDA$:
\begin{align*}
  \AdvINDCDDA_{\EDB}[(\adv)] =& |\Pr[\INDCDDA^{\adv}_{1,q,S,f} \Rightarrow 1] - \\
                           & \Pr[\INDCDDA^{\adv}_{0,q,S,f} \Rightarrow 1]|.\\
\end{align*}
\noindent
The following theorem establishes the security of \name under $\INDCDDA$:
\begin{theorem}[\INDCDDA\xspace Security]
  \label{theorem:indcdda}
  Let $q \geq 0$ and $Q = q \cdot B$. Let $\dist_0, \estdist_0, \dist'_0, \estdist'_0, \dist_1, \estdist_1, \dist'_1, \estdist'_1$ be query distributions.
  For any $q$-query $\INDCDDA$ adversary $\adv$ against $\name$ there exist adversaries $\bdv$, $\cdv$, $\ddv_1$, $\ddv_2$, $\ddv_3$,$\ddv_4$ such that \\
  $$\AdvINDCDDA_{\name}[(\adv)] \leq \AdvPRF_{F}[(\bdv)] + \AdvROR_{E}[(\cdv)]$$
  $$\myind\myind\myind\myind\myind\myind\myind\myind\myind + \AdvDIST_{Q,\dist_0,\estdist_0}[(\ddv_1)] + \AdvDIST_{Q,\dist'_0,\estdist'_0}[(\ddv_2)]$$
    $$\myind\myind\myind\myind\myind\myind\myind\myind\myind + \AdvDIST_{Q,\dist_1,\estdist_1}[(\ddv_3)] + \AdvDIST_{Q,\dist'_1,\estdist'_1}[(\ddv_4)]$$
  where $F$ and $E$ are the $PRF$ and $AE$ schemes used by \name. Adversaries
  $\bdv, \cdv, \ddv_1,\ddv_2,\ddv_3,\ddv_4$ each run in same time as $\adv$, and make $Q$ queries.
\end{theorem}

\begin{proof}
  We first give a proof sketch, followed by the details.
  Our proof is similar to that for $\INDCDA$ security --- we show that the transcripts generated by \name both before and after the distribution change are independent of the underlying distributions. Clearly, $\INDCDA$ security already guarantees that the output transcripts before the distribution changes and after replica swapping phase completes in \name are independent of the underlying distribution. The interim period comprises detecting the distribution change and performing replica swaps in response. Our distribution change detection occurs at a single leader \li proxy server that observes keys of all client queries (\S\ref{ssec:three-layer}, \S\ref{ssec:dynamic_dist} in the main paper), and employs the same detection mechanism as \pancake. As such, the scheme ensures that the output transcript is independent of the underlying distribution in the interim phase, provided the distribution change occurs instantaneously.
  If the detection takes finite time, \name --- as well as \pancake --- experience a short period of bias in distribution proportional to the difference between $\estdist$ and $\newestdist$. However, prior work~\cite{pancake} shows that exploiting this bias to learn anything meaningful about the underlying distribution is challenging. 
  We now detail our formal proof, leveraging the security of \pancake to ensure output transcripts are independent of the access distributions $\dist$, $\newdist$.
  
  \paragraph{Proof Details.} Similar to the proof for $\INDCDA$ security, we substitute $\Init$ and $\Process$ in $\INDCDDA$ game with the algorithms in \name. Next, we replace the PRF $F$ and the encryption scheme $E$ with a truly random function, and replace $\estdist_0,\estdist'_0,\estdist_1$ and $\estdist'_1$ with $\dist_0,\dist'_0,\dist_1$ and $\dist'_1$ respectively to obtain game $G$. Again, the difference between the success of an adversary $\advA$ in the game $G$ and $\INDCDA$ can be upper bounded by the sum of advantages of (i) the PRF adversary $\advB$ against the PRF scheme $F$, (ii) the real-or-random adversary $\advC$ against the encryption scheme $E$, (iii) the distribution adversaries $\advD_1$, $\advD_2$, $\advD_3$ and $\advD_4$  in computationally distinguishing between distributions $\estdist_0 \leftrightarrow \dist_0$, $\newestdist_0 \leftrightarrow \newdist_0$, $\estdist_1 \leftrightarrow \dist_1$, and $\newestdist_1 \leftrightarrow \newdist_1$, respectively:
\begin{align}
  \label{eq:diff_adv2}
  |\underset{b \sim \{0,1\}}{\Pr}[&\INDCDDA^{\adv}_{b,q,S,f} \Rightarrow b] - {\Pr}[G^{\adv}_{b,q,S,f} \Rightarrow b]| \nonumber \\
                                  &\leq \AdvPRF_{F}[(\bdv)] + \AdvROR_{E}[(\cdv)] + \AdvDIST_{Q,\dist_0,\estdist_0}[(\ddv_1)] \nonumber \\
                                  & + \AdvDIST_{Q,\dist'_0,\estdist'_0}[(\ddv_2)] + \AdvDIST_{Q,\dist_1,\estdist_1}[(\ddv_3)] \nonumber \\
                                  & + \AdvDIST_{Q,\dist'_1,\estdist'_1}[(\ddv_4)]
\end{align}
\noindent
We now argue that the underlying distribution is indistinguishable to the adversary. Formally,
\begin{align}
  \label{eq:ind_g2}
  \underset{b \sim \{0,1\}}{\Pr}[G^{\adv}_{b,q,S,f} \Rightarrow b] = 1/2 
\end{align}
\noindent
In other words,
\begin{align}
  {\Pr}[G^{\adv}_{0,q,S,f} \Rightarrow 1] = {\Pr}[G^{\adv}_{1,q,S,f} \Rightarrow 1] \nonumber
\end{align}

\begin{figure}
  \centering
  \iffullversion
  \includegraphics[width=\columnwidth]{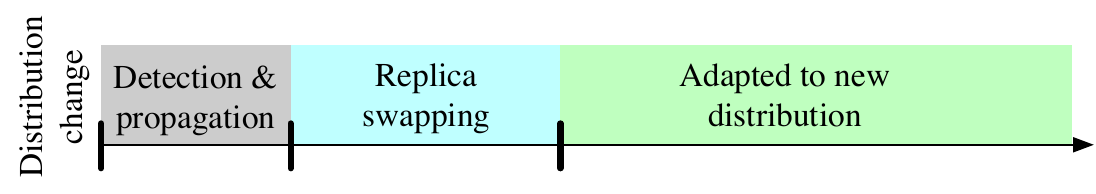}
  \else
  \includegraphics[width=\columnwidth]{../img/distchange}
  \fi
  \caption{\textbf{Three distinct phases in \name after distribution change.} See proof of $\INDCDDA$ security for details.}
  \label{fig:distchange}
\end{figure}
\noindent
$\tau_{1}, \tau_{2}, ..., \tau_{|C_{\liii}|}$ are already independent of the underlying distribution $\dist$ due to $\INDCDA$ security. For $\tau'_{1}, \tau'_{2}, ..., \tau'_{|C_{\liii}|}$ transcripts, we consider three distinct phases post distribution change, as shown in Figure~\ref{fig:distchange}: detection and propagation, replica swapping, and post-adaptation. For the subset of transcripts $\tau'_{1}, \tau'_{2}, ..., \tau'_{|C_{\liii}|}$ that lie in post-adaptation phase, independence from the underlying distribution also follows from $\INDCDA$ security. For subset of transcripts $\tau'_{1}, \tau'_{2}, ..., \tau'_{|C_{\liii}|}$ that lie in the detection and propagation and replica swapping phases, we leverage the security provided by the oblivious data access scheme $\ninj$, and the distribution change atomicity afforded by the atomic replica initiation protocol. In particular, $\ninj$ ensures that as long as the distribution change is atomic and instantaneous (as discussed above), the output transcript during the detection and propagation and replica swapping phases is independent of the underlying distribution. At the same time, distribution change atomicity ensures the existence of an instant of time $t_{change}$ when \name switches from processing queries according to $\estdist$ to processing them according to $\newestdist$ (\S\ref{ssec:dynamic_dist} of the main paper). Combining the above arguments shows that the transcripts $\tau'_{1}, \tau'_{2}, ..., \tau'_{|C_{\liii}|}$ are indeed independent of the underlying distribution $\newdist$. Therefore, an adversary cannot use the output transcripts $\tau_{1}, \tau_{2}, ..., \tau_{|C_{\liii}|}$ or $\tau'_{1}, \tau'_{2}, ..., \tau'_{|C_{\liii}|}$ to distinguish between two input distributions to the $\INDCDDA$ game, completing our proof.  
\end{proof}


  \iffullversion
\else
\customlabel{inv:distchange_atomicity}{1}
\fi
\section{Distribution Change Atomicity Proof}
\label{app:2pc}

\begin{figure}[t]
  \centering
  {\setlength{\fboxsep}{.05\fboxsep}
    \framebox{
      \begin{tabular}{l}
      \begin{minipage}[t][4.5em]{.45\textwidth}\gamesfontsize\setstretch{1.1}
        \procedurev{\textbf{Messages}}\\
        \texttt{prepare}, \texttt{prepareACK}: Marks beginning and end of prepare phase\\
        \texttt{commit}, \texttt{commitACK}: Marks beginning and end of commit phase\\
        \texttt{flushEOF}: Acknowledges all pending queries are flushed
      \end{minipage}
      \end{tabular}
    }
    \framebox{
      \begin{tabular}{l}
      \begin{minipage}[t][11em]{.215\textwidth}\gamesfontsize\setstretch{1.1}
        \procedurev{\textbf{\li Leader}}\\
        \underline{\textit{On distribution change}:}\\
        \ind Send \texttt{prepare} to all \li, \lii, \liii\\
        \ind Wait for all \texttt{prepareACK}s\\\vspace{-0.75em}\\
        \underline{\textit{On receiving all} \texttt{prepareACK}s:}\\
        \ind Send \texttt{commit} to all \li, \lii, \liii\\
        \ind Wait for all \texttt{commitACK}s\\\vspace{-0.75em}\\
        \underline{\textit{On receiving all} \texttt{commitACK}s:}\\
        \ind End
      \end{minipage}
      \end{tabular}
    }%
    \framebox{
      \begin{tabular}{l}
      \begin{minipage}[t][11em]{.205\textwidth}\gamesfontsize\setstretch{1.1}
        \procedurev{\textbf{\li server}}\\
        \underline{\textit{On} \texttt{prepare}:}\\
        \ind Start queueing client queries\\
        \ind Flush pending queries to \lii\\
        \ind Send \texttt{flushEOF} to all \lii\\
        \ind Send \texttt{prepareACK} to leader\\\vspace{-0.75em}\\
        \underline{\textit{On} \texttt{commit}:}\\
        \ind Stop queueing client queries\\
        \ind Send \texttt{commitACK} to leader
      \end{minipage}
      \end{tabular}
    }
    \framebox{
      \begin{tabular}{l}
      \begin{minipage}[t][9.25em]{.21\textwidth}\gamesfontsize\setstretch{1.1}
        \procedurev{\textbf{\lii server}}\\
        \underline{\textit{On} \texttt{prepare}:}\\
        \ind Wait for \texttt{flushEOF} from all \li\\
        \ind Flush pending queries to \liii\\
        \ind Send \texttt{flushEOF} to all \liii\\
        \ind Send \texttt{prepareACK} to leader\\\vspace{-0.75em}\\
        \underline{\textit{On} \texttt{commit}:}\\
        \ind Send \texttt{commitACK} to leader\\
      \end{minipage}
      \end{tabular}
    }%
    \framebox{
      \begin{tabular}{l}
      \begin{minipage}[t][9.25em]{.21\textwidth}\gamesfontsize\setstretch{1.1}
        \procedurev{\textbf{\liii server}}\\
        \underline{\textit{On} \texttt{prepare}:}\\
        \ind Wait for \texttt{flushEOF} from all \lii\\
        \ind Flush pending queries to $\DB'$\\
        \ind Send \texttt{prepareACK} to leader\\\vspace{-0.75em}\\
        \underline{\textit{On} \texttt{commit}:}\\
        \ind Send \texttt{commitACK} to leader\\
      \end{minipage}
      \end{tabular}
    }
  }\vspace*{-.25em}
  \caption{\textbf{Atomic initiation protocol for replica-swapping.}}
  \label{fig:replicaswap_init}
  \vspace{-1.5em}
\end{figure}

We now prove that our 2-PC inspired protocol for atomic initiation of replica swapping guarantees the Distribution Change Atomocity Invariant. In order to prove this, we first precisely specify the details of our protocol. Note that the termination phase after replica swapping employs a similar approach. The protocol is depicted in Figure~\ref{fig:replicaswap_init}:

\begin{itemize}[leftmargin=*, itemsep=0pt]
  \item  On detecting a distribution change from $\estdist$ to $\newestdist$, the leader sends 
 \texttt{prepare} messages to all the \li, \lii and \liii servers.
  \item  On receiving a \texttt{prepare} 
 message, \li servers temporarily stop processing new queries and start queueing them up --- these 
 correspond to the new distribution $\newestdist$. They subsequently flush pending queries from $\estdist$ 
 that have not yet been sent to \lii proxy servers. Once all such queries are flushed, they send a \texttt{flushEOF} 
 message to all the \lii proxy servers, and a \texttt{prepareACK} message to the leader.
  \item  \lii servers, on receiving 
 a \texttt{prepare} message, first wait till they have received \texttt{flushEOF} from \emph{all} \li proxy servers,
 and then flush all pending queries to the \liii servers. Similar to \li servers, they then send \texttt{flushEOF} 
 to all the \liii servers and respond to the \li leader with a \texttt{prepareACK} message.
  \item   \liii servers, upon receiving a \texttt{prepare} message, take actions similar to \lii servers: wait for all \texttt{flushEOF} 
 messages \lii servers, flush all pending queries to the KV store, and respond to the \li leader with a 
 \texttt{prepareACK} message.
  \item  Once the \li leader receives \texttt{prepareACK} from all other servers, 
 it sends a \texttt{commit} message to all of them. 
  \item  On receiving \texttt{commit}, the \li servers 
 stop queueing client queries and start processing them again according to $\newestdist$, as per
 the replica-swapping approach employed by noise-injection schemes.
\end{itemize}

Similar to replica swapping in the single-proxy approach, for every key that gains a replica, another key must lose one. The leader proxy determines the pairs of replicas that need to be swapped, the temporary fake access distribution and other relevant information. \name communicates this information to the other \li and \lii proxy servers by piggybacking it onto \texttt{prepare} messages. More precisely, the \li leader proxy computes and embeds the following information in \texttt{prepare} messages:
\begin{itemize}[leftmargin=*, itemsep=0pt]
    \item The new real distribution, new fake distribution, and temporary fake distribution used during replica swapping.
    \item The list of keys that are to gain and lose replicas, along with the mapping between them.
    \item The updated mapping from plaintext keys to their replicas.
\end{itemize}
\noindent
Since the protocol for initiating and terminating the replica-swapping phase takes finite time, an adversary may be able to detect that a distribution change has occurred. However, we show in \S\ref{sec:security} of the main paper that the atomicity of replica-swapping initiation and termination, coupled with the security of replica swapping in noise-injection schemes, ensures that the adversary cannot glean any information about $\estdist$ or $\newestdist$ under \name.

We now prove that the above protocol guarantees the distribution change atomicity invariant:
\begin{invariant}[Distribution change atomicity]\label{inv:distchange_atomicity}
  Once any \liii proxy server issues a query according to $\newestdist$, all subsequent queries issued by any \liii server must be according to $\newestdist$.
\end{invariant}
\begin{proof}
  We prove that the invariant holds for a single distribution change event from $\estdist$ to $\newestdist$. It can easily be generalized to an arbitrary sequence of distribution change events using induction.
  
  To prove this we show that there exists an instant of time $t_{change}$ in the protocol's execution, such that: (1) before $t_{change}$, all queries are processed according to the distribution $\estdist$, and (2) after $t_{change}$, all queries are processed using $\newestdist$. We show that $t^* = $ the time instant when the \li leader receives \texttt{prepareACK} from all other servers, satisfies this property.

  First, we prove that a query processed according to $\newestdist$ can only be forwarded by an \liii server after $t^*$. The \li servers start processing queries according to $\newestdist$ only after they receive a \texttt{commit} message. Such a \texttt{commit} message can only be delivered once it has been sent by the \li leader. A \texttt{commit} can only be sent by the \li leader after $t^*$. Hence, an \li server can only start processing queries according to $\newestdist$ after $t^*$. A query can be issued by an \liii server only after it has been processed at an \li server. Hence, a query processed according to $\newestdist$ can only be forwarded by an \liii server after $t^*$.

  Second, we prove that queries according to $\estdist$ cannot be forwarded by an \liii server on or after $t^*$. Before $t^*$ (when \li leader has received all \texttt{prepareACKs}), all of the \liii servers would have flushed any outstanding queries, before which all \lii servers would have flushed their outstanding queries, before which all \liii servers would have flushed their outstanding queries. Hence, it is guaranteed that by $t^*$ all queries processed according to $\estdist$ have been issued by the \liii servers.

  Combining the above two facts implies that all queries issued by \liii servers after $t^*$ are those processed according to $\newestdist$, while all queries before $t^*$ are those processed according to $\estdist$, hence proving the invariant. 
\end{proof}
\else
  {
  \balance
  \bibliographystyle{unsrt}
  \bibliography{bib/citations}
  \appendix
  }
\fi
\end{sloppypar}

\end{document}